\documentclass[%
preprint,
nofootinbib,
 amsmath,amssymb,
 aps,
floatfix,
]{revtex4-2}

\usepackage[utf8]{inputenc}
\usepackage{graphicx}
\usepackage{dcolumn}
\usepackage{bm}
\usepackage{hyperref}
\usepackage[dvipsnames]{xcolor}
\usepackage{makecell}

\usepackage{fancyhdr}
\fancypagestyle{plain}{ %
  \fancyhf{} 
  
}
\fancypagestyle{titleStyle}
{
   \fancyhf{}
   
   \fancyhead[L]{153-120000-CONF-001967-REV 0.1}
   \fancyhead[R]{UNRESTRICTED}
   \fancyfoot[L]{\textcopyright Canadian Nuclear Laboratories}
}

\begin{document}

\title{Neutrino-based safeguards of CANDU spent fuel using superconducting detectors and the CE$\nu$NS interaction}
\author{M. Stringer}
\email{mark.stringer@cnl.ca}
\author{A. Erlandson}
\author{V. Anghel}  
\author{Z. Yamani} 
\affiliation{Canadian Nuclear Laboratories\\
286 Plant Road \\
Chalk River, ON K0J 1J0 \\
Canada}

\begin{abstract}
    To prevent the unauthorised spread of radioactive materials, it is essential to detect and monitor spent nuclear fuel. This paper investigates the feasibility of using a detector based on transition edge superconductors to monitor spent CANDU fuel in two distinct scenarios. The first of these considered the monitoring of a CANSTOR container at the location of the Gentilly-2 nuclear power plant. The fuel in a CANSTOR container is contained within baskets, which are then stored in tubes in the container. An individual detector with a mass of 1~kg  is not sensitive to the removal of a single basket from the container without an unfeasibly long monitoring time. When an entire tube in the container is emptied (equivalent to approximately 5\% of the fuel in the container), the results are improved.  The second scenario examined  the feasibility of monitoring a single dry storage container (DSC) at the Pickering site. The DSCs are much smaller than a CANSTOR container and contain approximately 7.4 tonnes of spent fuel. The background due to the neutrinos from the nearby reactors at the Pickering site was also evaluated. It was found that monitoring DSC was unfeasible due to the high reactor neutrino background. Both studies utilized fuel that was 30 years old; monitoring a container loaded with spent fuel that is younger would require a shorter monitoring time due to the fuel's higher radioactivity.
\end{abstract}

\date{\today}
\maketitle
\thispagestyle{titleStyle}

\section{Introduction}
The monitoring and accountancy of spent nuclear fuel are crucial to prevent the proliferation of nuclear materials. 

Neutrinos are a candidate of interest to monitor spent fuel~\cite{2021arXiv211212593A}~\cite{PhysRevD.105.056002}. Their minuscule cross-section means that any detected neutrinos will be unaffected by any of the shielding surrounding the spent fuel, and that the rate of events will simply depend on the distance between the spent fuel and the detector. Furthermore the shielding will not alter the energy spectra of the neutrinos in any way. 
However, the tiny interaction cross-section of the neutrinos poses several experimental challenges as the low interaction probability also results in a low event rate.

Recent studies have looked into placing a large liquid scintillator-based detector near the spent fuel to detect $\bar{\nu}_e$ produced in the spent fuel via the inverse beta decay (IBD) interaction~\cite{PhysRevApplied.8.054050}. The scintillator detectors would need to be on the order of tens of tonnes in order to get a sufficient event rate. The large size of these detectors and the large amount of equipment associated with them may make their placement at a spent fuel storage site difficult in practice. 

Detectors based on Transition Edge Sensors (TESs)~\cite{Irwin2005} use a much smaller active detector volume and therefore are much more compact than traditional scintillator detectors. The size of the detector can be much smaller than traditional detectors by exploiting the recently observed coherent elastic neutrino-nucleus scattering (CE$\nu$NS)~\cite{Scholberg:2018vwg}\cite{doi:10.1126/science.aao0990}\cite{PhysRevLett.126.012002}, which has a much larger cross-section than both electron scattering and inverse beta decay.  

This study describes the detection method of TES-based neutrino detectors and calculates the anticipated signal from spent CANDU\textsuperscript{\textregistered}\footnote{CANDU is a registered trademark of Atomic Energy of Canada Ltd.}(Canada Deuterium Uranium) fuel~\cite{CANDUFuel} stored in waste fuel containers. Possible signal backgrounds are investigated, and the viability of using the technique to monitor spent CANDU fuel is given. We explore the interaction rates of Calcium, Tungsten and Oxygen as they are the constituent elements of the target (CaWO$_{4}$) planned for use in the NUCLEUS experiment~\cite{Wagner_2021}. The successful operation of the NUCLEUS detector will confirm the viability of this technique for monitoring spent fuel.   

The fuel from CANDU reactors is stored in different containers depending on the reactor site in question. The spent fuel produced at Gentilly-2 nuclear power plant is stored in large concrete CANSTOR containers~\cite{CANSTORInfo} each container can store up to 12,000 spent fuel bundles. Nuclear fuel produced at the Pickering and other plants in Ontario is stored in smaller containers known as dry storage containers (DSCs); each DSC can store up to 96 fuel bundles~\cite{DSCInfo}. The background event rate at spent fuel storage sites is highly unknown and highly dependent on detector design. Also, rates will vary depending on location; for example, the DSC at Pickering will have an additional background due to neutrinos produced in the adjacent operating reactors.
In this paper we first describe the  mechanism for the detection of neutrinos using a TES. We then describe how the expected rate of events due to neutrinos emitted by the spent fuel is calculated, before summarising the expected backgrounds to the neutrino signal. The sensitivity of the detector to the removal of spent fuel from both CANSTOR and DSC detectors is presented followed by a discussion of the results and the feasibility of the technology to monitor spent fuel. 
 
\section{TES-Based Neutrino Detection}
TES-based neutrino detection uses a TES thermally coupled to a cryogenically cooled bolometer. When a particle interacts within the bolometer it produces a miniscule rise in the bolometer's temperature, which is detected by the TES.
\subsection{The CE$\nu$NS Interaction}
The interaction channel used to detect neutrinos is coherent elastic neutrino nucleus scattering (CE$\nu$NS). In this interaction the neutrino interacts with the nucleus as a whole, scattering elastically. The interaction is flavour insensitive, which means this interaction provides a value of the flux of neutrinos of all flavours rather than just one.
The differential cross-section of the interaction is given by~\cite{PhysRevD.94.063527}:
\begin{equation}
    \frac{d\sigma}{dE_r}(E_\nu, E_r) = \frac{G^2_F}{4\pi}Q_W^2 m_N \left(1-\frac{m_N E_r}{2E^2_\nu}\right) F^2(E_r)
\end{equation}
\begin{equation}
    Q_W = N-(1-4\textrm{sin}^2\theta_W)Z
    \label{eq:qw}
\end{equation}
where $G_F$ is the Fermi coupling constant, $m_N$ is the mass of the nucleus, $Q_W$ represents the weak nuclear charge, $F(E_R)$ represents the form factor, $E_r$ is the recoil energy of the nucleus and $E_\nu$ is the energy of the incoming neutrino. In Eq.~\ref{eq:qw} $N$ is the number of neutrons within the nucleus and $Z$ the number of protons, and $\sin^2\theta_W\approx0.231$.

The value used in the following sections for the form factor is~\cite{PhysRevD.82.023530}: 

\begin{equation}
F^2(E_r) = \left(\frac{3j_1(qR)}{qR}\right)^2e^{-q^2s^2}
\end{equation}
with $q=\sqrt{2m_N E_r}$ and $R=\sqrt{c^2+\frac{7}{3}\pi^2a^2-5s^2}$. The value of c is dependent on the mass number ($A$) of the nucleus and is given by $c=1.23A^{1/3}-0.6$~fm; the values for $s$ and $a$ are 0.9~fm and 0.52~fm respectively.

The approximate maximum recoil energy of the interaction is given by $2E_\nu^2/m_N$.

The dependence of the cross-section on the product of $m_N$ and $N$ means many more interactions will take place with a heavier nucleus. However, the kinematics of the interactions causes heavier nuclei to recoil with a lower kinetic energy. Hence, a lower detector threshold is required to fully exploit this increased interaction rate when a heavier target nucleus is used. 
\vspace*{5 mm} 
\subsection{The TES mechanism}
The interaction of the particle within the target results in a miniscule rise in the temperature of the target. The rise in temperature depends on the choice of material and the temperature of the target~\cite{pyle2015optimized}. Using the 300~g CaWO$_{4}$ crystal described in~\cite{pyle2015optimized} and approximating the temperature increase simply as $\Delta T=\Delta E/C$ where $C$ is the heat capacity of the crystal a 500~keV deposition of energy within the target results in a 0.61~$\mu$K increase in the crystals temperature. Therefore, in order to detect these interactions the temperature of the target must be measured with accuracy on the order of $\mu$K. TESs act as very sensitive thermometers and measure this change in temperature~\cite{Irwin2005}. They consist of a piece of superconducting material kept just below the critical temperature of the material. When the TES is heated above its critical temperature, it ceases to be superconducting and its resistance begins to increase from zero, resulting in a minor potential difference across the TES. This minor voltage differential can be read and converted into an electronic signal for the interaction.
Because the TES operates so close to the critical temperature, it must be kept in a temperature-controlled environment. Instead of directly detecting the voltage, a superconducting quantum interference device (SQUID) can be used as a magnetometer to measure the change in current across the TES as a result of heating above its critical temperature. Utilizing the SQUID in this manner reduces the number of thermal couplings between the TES and other components, as well as the detector's thermal noise.
 The fundamental threshold of the detector depends on the thermal noise of the detector~\cite{PhysRevD.85.013009}. The noise from the target depends on the target mass, material and operating temperature; increasing the size of the target increases the noise and threshold of the detector, although this can be compensated for by decreasing the temperature~\cite{PhysRevD.85.013009}. An alternative option is to use multiple target crystals with masses of the order g with a TES monitoring each individual target, this is the method that the NUCLEUS experiment plans to use to scale up to a 1~kg detector mass~\cite{2017EPJC...77..506S}. Note that the critical temperature of the TES needs to be tuned to the operating temperature of the cryostat.   
\section{Determining the Rate of Neutrino Events}
The determination of the neutrino interaction rate can be broadly split into two parts. The first of these is the determination of the neutrino flux. This involves finding the composition of the spent fuel in records and then combining the composition with the neutrino emission spectrum and activities of each of the isotopes in order to construct the overall spectrum. The second part of the calculation involves scaling the neutrino spectrum to account for the distribution of the spent fuel in the container, and then combining it with the interaction cross-sections to obtain an event rate. 

\subsection{Construction of the neutrino spectrum from spent CANDU fuel}
To construct the overall neutrino spectrum from the spent fuel, the composition of the fuel and the neutrino spectra from the constituent isotopes are combined. The composition of the fuel and the Zircaloy cladding was obtained from~\cite{NWMOSpentFuelComp}. The neutrino spectra were obtained by using the BetaShape spectra~\cite{TabRad_v1,TabRad_v2,TabRad_v3,TabRad_v4,TabRad_v5,TabRad_v6,TabRad_v7,TabRad_v8}. Several of the listed isotopes in the spent fuel lacked BetaShape predictions, and therefore they were not included in the flux; this resulted a 0.1\% reduction in the total neutrino flux.

\begin{figure*}
    \centering
    \includegraphics[width=0.8\linewidth]{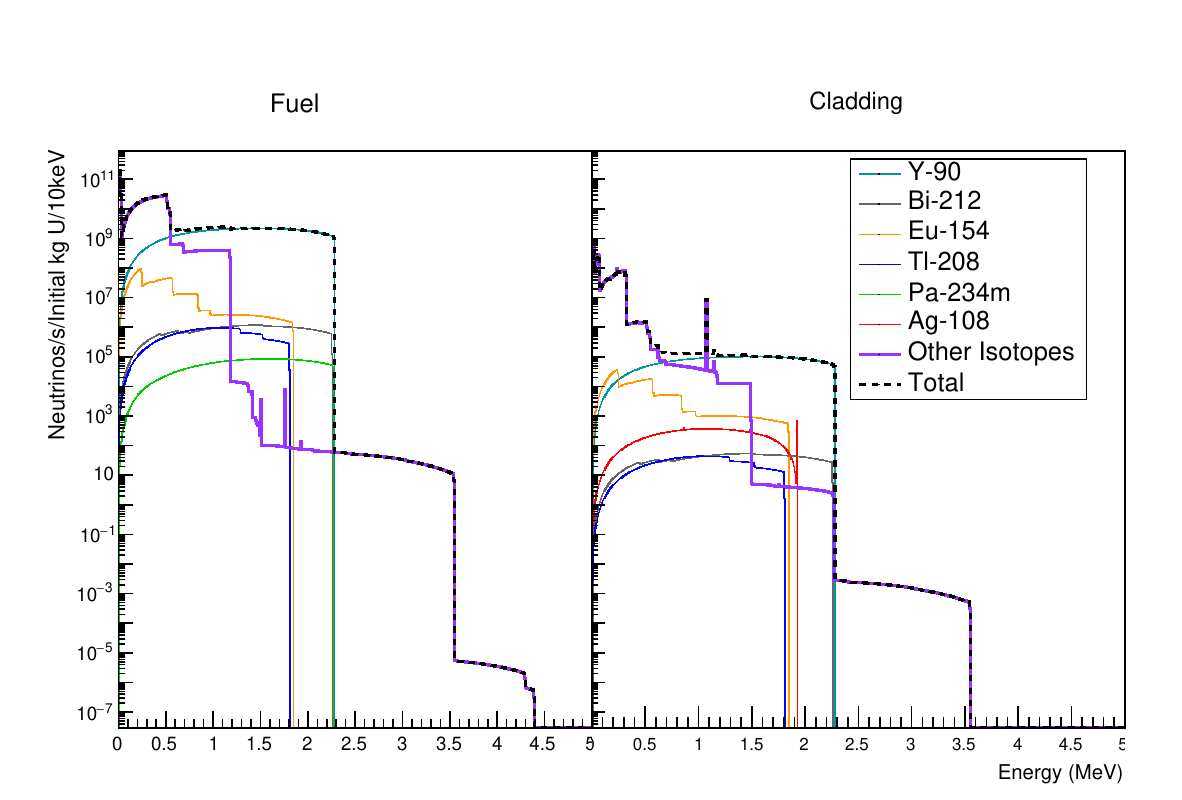}
    \caption{The neutrino spectra from spent fuel and cladding 30 years after removal from the reactor calculated using the fuel composition in~\cite{NWMOSpentFuelComp} and the BetaShape spectra.}
    \label{fig:fuelEmissions}
\end{figure*}

The neutrino spectra from the fuel and the cladding 30 years after being removed from the reactor calculated using this method is shown in Figure~\ref{fig:fuelEmissions}. The five isotopes with the largest neutrino flux above 1.5 MeV are shown as well as the sum of the other isotopes. Similar to other spent fuel the dominant neutrino emitter is found to be $^{90}$Y~\cite{PhysRevApplied.8.054050}. The remainder of the named isotopes in the figure contribute only a small fraction to the overall neutrino rate. Past the endpoint of the $^{90}$Y the neutrino flux drops by eight orders of magnitude. Neutrinos are produced up to beyond 4~MeV; however, at that point the flux of neutrinos is less than a factor of $10^{-10}$ of the peak flux.  The neutrino emissions from the Zircaloy cladding surrounding the fuel are also dominated by the emissions from $^{90}$Y. The overall flux from the cladding is small compared to that of the fuel, and therefore only the emissions from the fuel are studied in the following sections.

\subsection{Geometric effects}
\begin{figure}
    \centering
    \includegraphics[width=\linewidth]{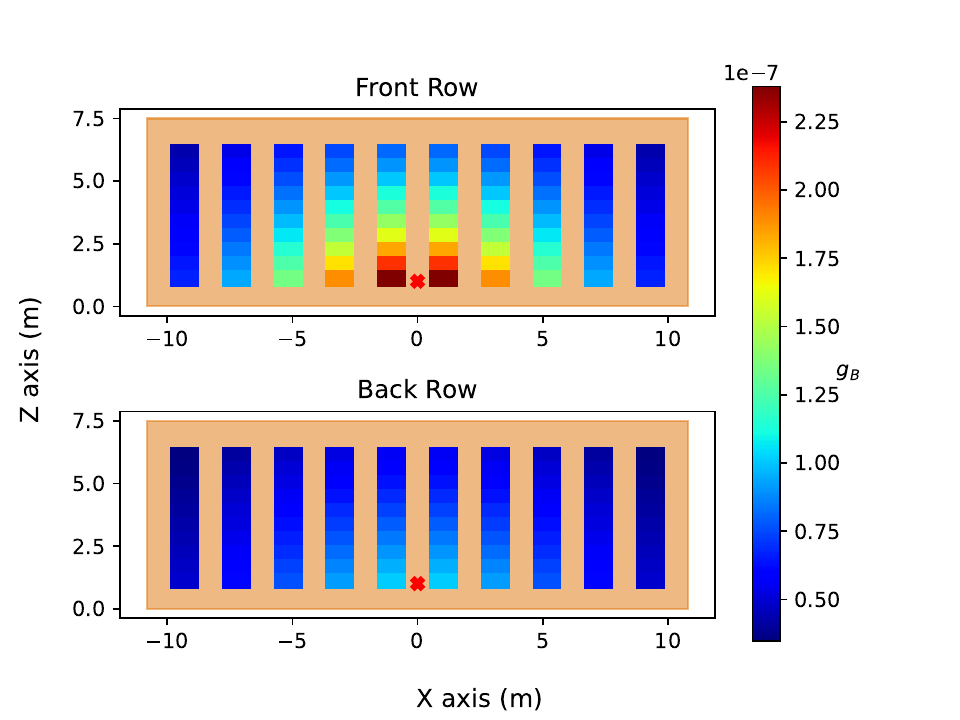}
    \caption{The geometric factor $g_B$ in units of cm$^{-2}$ for each of the baskets in a CANSTOR container when viewed from the front. The detector is positioned 2~m from the center of the long face of the container and 1~m off of the ground.}
    \label{fig:geometricFactorFront}
\end{figure}

A CANSTOR container is 21.6 m long, 8.1 m high and 7.5 m deep. Twenty tubes are distributed in two rows of ten, with the spent fuel stored in cylindrical steel baskets in the tubes~\cite{CANSTORInfo}. Each basket contains 60 fuel bundles and each tube contains ten baskets. The positions of the tubes within the container were approximated based on the drawings in~\cite{CANSTORInfo}.  Figure~\ref{fig:geometricFactorFront} shows the geometric factor $g_B$ (used later in Eq.~\ref{eq:overallRateAllBaskets}). In order to evaluate this, all the fuel in each basket and the detector were treated as a point source, giving the simple scale factor $1/(4\pi d^2)$ where $d$ is the distance between the center of the fuel basket and the center of the target material. The fuel in the tubes farther from the sensors contributes significantly less than the nearby fuel.

\subsection{The overall event rate}

The overall event rate assuming a point-like source and detector is given by Eq.~\ref{eq:pointSourceRate}, where $g$ is the geometric factor (given by $1/(4\pi d^2)$ where $d$ is the distance between the detector and source), $N_T$ is the number of target nuclei within the detector, $\frac{d\sigma(E_\nu)}{dT}$ is the differential cross-section given by Eq.~\ref{eq:qw} and $f(E_\nu)$ is the flux of neutrinos from the spent fuel.

\begin{equation}
    \frac{dR}{dT} = g N_T  \int \frac{d\sigma(E_\nu)}{dT}f(E_\nu) dE_\nu
    \label{eq:pointSourceRate}
\end{equation}

In the following studies each fuel basket is considered to be a point source. Assuming all the fuel in the container has the same age, the overall observed flux from the CANSTOR container will be the sum of these point sources:

\begin{equation}
     \frac{dR_T}{dT} = \left( N_T  \int \frac{d\sigma(E_\nu)}{dT}f(E_\nu) dE_\nu \right) \sum_B g_B
     \label{eq:overallRateAllBaskets}
\end{equation}
where $g_B$ is the geometric factor for each basket.

For the monitoring of the DSC the inner cavity of the container is considered to be a uniform source of neutrinos. The DSC can be approximated as a point source where the geometric factor $g$ in Eq.~\ref{eq:pointSourceRate} is calculated using:
\begin{equation}
    g = \frac{1}{4\pi} \int\int\int \frac{1}{|\vec{x}-\vec{p}|^2} dx dy dz \times \frac{1}{V}
    \label{eq:geometricFactorDSC}
\end{equation}
where $\vec{x} =(x,y,z)$ is the vector being integrated over the inner cavity of the DSC, $\vec{p}$ is the position of the sensor and $V$ is the volume of the inner cavity. The dimensions of the inner cavity are (1.046, 1.322, 2.520) m~\cite{DSCDimensions}. Assuming the sensor is placed 1~m off of the ground and 2~m from the center of the long side of the DSC, the value of $g$ is $7.746\times 10^{-7}$~cm$^{-2}$.

\begin{figure}
    \centering
    \includegraphics[width=\linewidth]{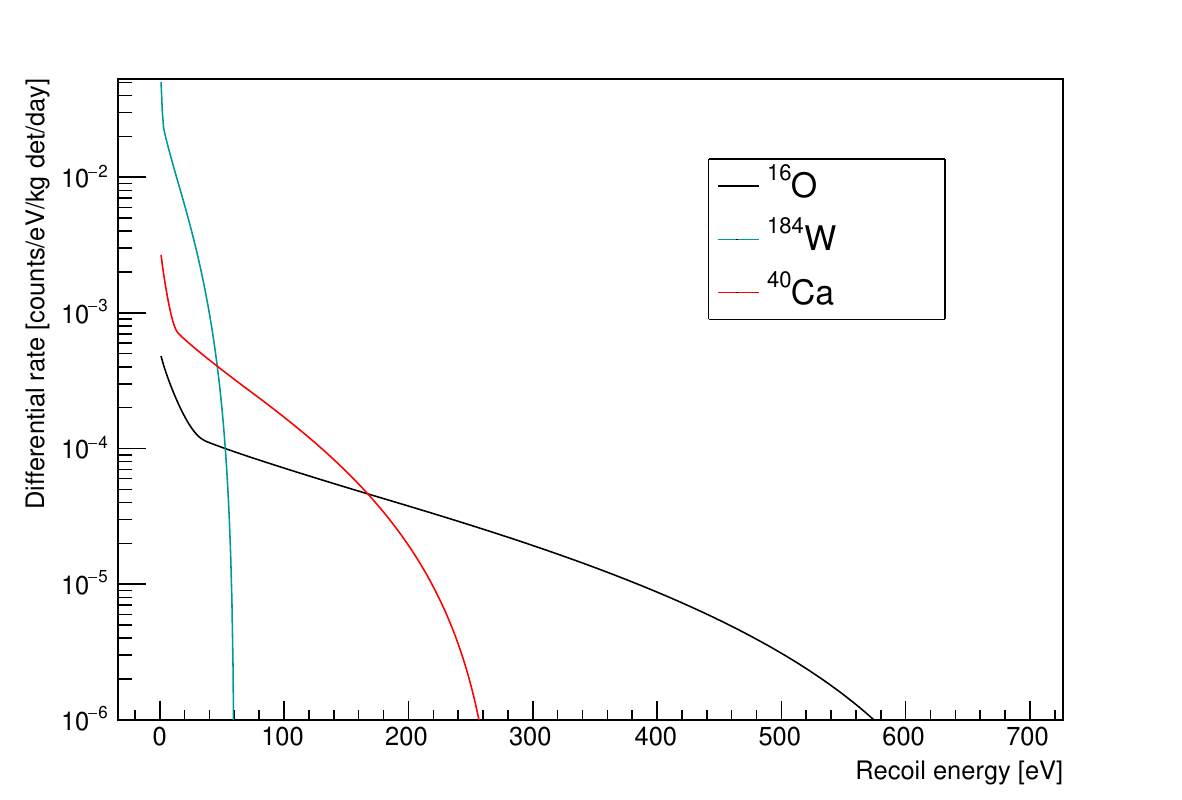}
    \caption{The recoil energy spectra of various target elements from 30-year-old spent fuel in a CANSTOR container.}
    \label{fig:diffInteractionRate}
\end{figure}

The $\frac{dR_T}{dT}$ distribution due to a full CANSTOR container for various target materials is shown in Figure~\ref{fig:diffInteractionRate}. The number of events detected depends on the threshold of the detector and the material. Heavier elements such as $^{184}$W have a much larger cross-section than lighter nuclei. However, due to their increased mass they recoil at a much lower energy than lighter nuclei such as $^{16}$O. The optimal choice of target material depends on the exact threshold achievable.

\section{Backgrounds to the Neutrino Signal From Spent Fuel}
Several sources of background could affect the rate of events, these include include secondaries due to cosmic ray muons and intrinsic gamma background events~\cite{Wagner_2021}. The rate of background events strongly depends on the location and design of the detector. Furthermore, the backgrounds in these types of detectors may not be fully characterised. Many dark matter experiments using a similar detection mechanism report an excess rate of events below 100~eV~\cite{Wagner_2021}. In the following section we make no assumption about the background rates and explore the same range of rates as in~\cite{PhysRevD.105.056002} from $10^{-2}$ background events per day up to 10,000 background events per day. A rate of 10,000 events is expected for a detector on the surface un-shielded from cosmic rays~\cite{PhysRevD.105.056002}. 

As spent fuel is often stored on site at reactor complexes, an additional source of background may be the neutrinos from the nearby reactors. As the Gentilly reactors are currently being decommissioned this source of background will not be present for monitoring of the CANSTOR containers. However, the monitoring of the DSC's at Pickering plant will be subject to this background. The background rate due to reactor neutrinos will depend on the position of the spent fuel relative to the reactors and the power output of the reactors. 

\subsection{Backgrounds due to nearby reactors at the Pickering Generating Station.}
At the time of drafting this manuscript six of the eight CANDU reactors at the Pickering plant are operational. In order to estimate the background due to the reactors the location of the monitoring of the DSCs is chosen to be the center of the ``Dry Storage Facility Phase 1" in Figure 1 of~\cite{DryStoragePickering} which shows a map of the Pickering site. The approximate distance between the monitoring area and each reactor was estimated using the same figure and the results are shown in Table~\ref{tab:reactorPowerAndDistances}. 

\begin{table}
    \centering
    \begin{tabular}{c|c}
         \textbf{Reactor Number} & \textbf{Distance to Reactor (m)} \\ \hline
         1 &  649.0\\
         4 &  448.8\\
         5 &  294.5\\
         6 &  239.7\\
         7 &  159.6\\
         8 &  112.5\\
    \end{tabular}
    \caption{The distance from the reactors at the Pickering site to the detector in the monitoring area, estimated using Figure 1 in~\cite{DryStoragePickering}.}
    \label{tab:reactorPowerAndDistances}
\end{table}

\begin{figure}
    \centering
    \includegraphics[width=\linewidth]{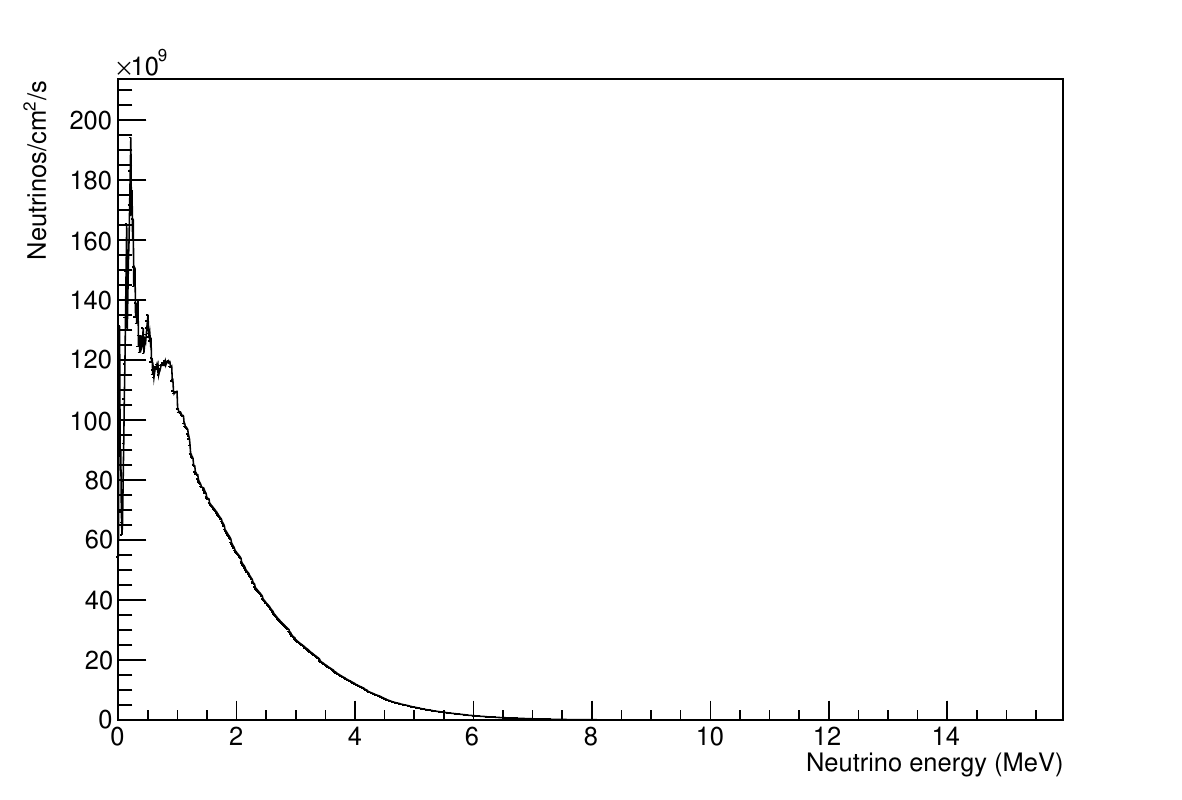}
    \caption{The calculated total neutrino flux from all the operating CANDU reactors calculated using the stand-off distances indicated in Table~\ref{tab:reactorPowerAndDistances} and the neutrino spectra in~\cite{canduReactorSpectrum}}.
    \label{fig:CANDUSpectra}
\end{figure}

The flux from the reactors at the dry storage facility can be seen in Figure~\ref{fig:CANDUSpectra}. The thermal power of all the cores was assumed to be 1,744~MW$_\textrm{th}$~\cite{PickeringPower}.

\begin{figure}
    \centering
    \includegraphics[width=\linewidth]{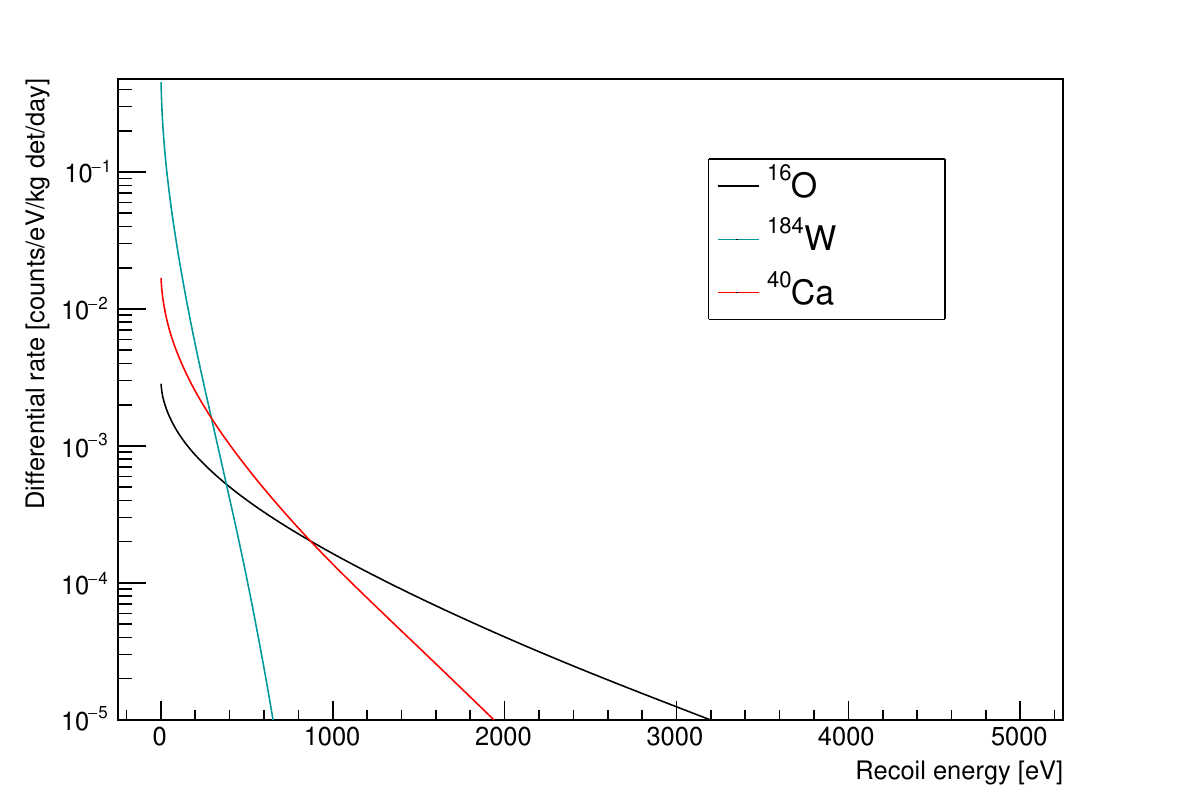}
    \caption{The calculated differential rate of recoil events for various targets due to neutrinos produced by the reactors at Pickering assuming the standoff distances in Table~\ref{tab:reactorPowerAndDistances}.}
    \label{fig:reactorDifferentialRate}
\end{figure}

The rate of events due to the reactors at Pickering  for various thresholds and targets is shown in Figure~\ref{fig:reactorDifferentialRate}. Due to the higher energies of the neutrinos the recoil energies are much higher than that of the spent fuel. For example, when the tungsten target is used the end point of the 30-year-old fuel is around 60~eV; in contrast, the end point for recoils due to neutrinos from the reactor is greater than 600~eV. 

The overall interaction rates for various thresholds are shown in Table~\ref{tab:rateThresholds}. As mentioned previously, the lighter nuclei show the lowest rates due to the smaller cross-section but are less affected by changes in threshold as their recoil energies are higher than that of heavier nuclei.

\begin{table}[b]
    \centering
    \begin{tabular}{c|c|c|c}
         Threshold (eV) & $^{40}$Ca & $^{184}$W & $^{16}$O  \\\hline
         1 &  1.73 & 11.65 & 0.70 \\
         5 & 1.67 & 10.18 & 0.69 \\
         10 & 1.61 & 8.83 & 0.68 \\
         20 & 1.50 & 6.94 & 0.65
    \end{tabular}
    \caption{The calculated interaction rate due to the reactor neutrino flux shown in Figure~\ref{fig:CANDUSpectra} in counts/kg det./day for various target isotopes and detector thresholds.}
    \label{tab:rateThresholds}
\end{table}

\section{Sensitivity of the Detector}

In this section, two scenarios are presented. The first is the monitoring of spent fuel stored in a CANSTOR container at the Gentilly-2 facility. The sensitivity of the detector to missing fuel at various locations in the container is investigated. The sensitivity to a missing single fuel basket and that to an entire tube being empty are investigated.
The second scenario involves the monitoring of waste fuel at the Pickering facility inside a single DSC. For a variety of thresholds and target materials, the exposure time necessary to distinguish between a full and partially filled container is investigated.

Regardless of the scenario the ability to distinguish between a full container and a container with missing fuel can be estimated by using the Gaussian approximation to the Poisson distribution as shown in Eq.~\ref{eq:differenceInRate}, where the term on the left represents the difference in counts between a full container and a container with missing fuel and the term on the right represents the uncertainty due to random fluctuations. $\mu_F$ is the rate of events due to a full container and $\mu_m$ is the rate of events due to a partially empty container. $t$ is the time for which the measurement takes place and $b$ is the rate of backgrounds.

\begin{equation}
    (\mu_F+b)t-(\mu_m+b)t = Z\sqrt{(\mu_F+b)t}
   \label{eq:differenceInRate}
\end{equation}

The exposure required to achieve a given sensitivity can be found by rearranging Eq.~\ref{eq:differenceInRate} for $t$, as shown in Eq.~\ref{eq:runTimeForConfidence}.

\begin{equation}
    t = \frac{Z^2(\mu_F+b)}{(\mu_F-\mu_m)^2}
    \label{eq:runTimeForConfidence}
\end{equation}

Rather than requiring a certain level of sensitivity, or confidence that the outcome of the measurement is not attributable to a statistical fluctuation, a limit on the rate of false alarms can be defined. A false alarm is defined as a measurement that falls outside of the confidence interval due to a random statistical fluctuation.

The time between false alarms $T$ can be defined as the measurement time divided by the false alarm probability, as shown in Eq.~\ref{eq:errorRate}.  

\begin{equation}
    T = \frac{t}{P(Z)}
    \label{eq:errorRate}
\end{equation}
where $P(Z)=\frac{1}{2}\left[1-\textrm{erf}\left(\frac{Z}{\sqrt{2}}\right)\right]$.

Substituting in the equations for $P(Z)$ and $t$ yields Eq.~\ref{eq:numericalTVsZ}.

\begin{equation}
    T = \frac{2(\mu_F+b)}{(\mu_F-\mu_m)^2}\frac{Z^2}{1-\textrm{erf}\left(\frac{Z}{\sqrt{2}}\right)}
    \label{eq:numericalTVsZ}
\end{equation}

Equation~\ref{eq:numericalTVsZ} can be numerically solved for $Z$, which in turn can be substituted into Eq.~\ref{eq:runTimeForConfidence} to determine the measurement time $t$ that would result in a given false alarm rate.

\subsection{Using the TES detector to monitor a CANSTOR container}
When the detector is used to monitor a CANSTOR container two scenarios have been studied. The first is that a single basket is missing from the container; the second is that an entire tube's worth of fuel missing.
\subsection{Removal of an individual basket}
\begin{figure}
    \centering
    \includegraphics[width=\linewidth]{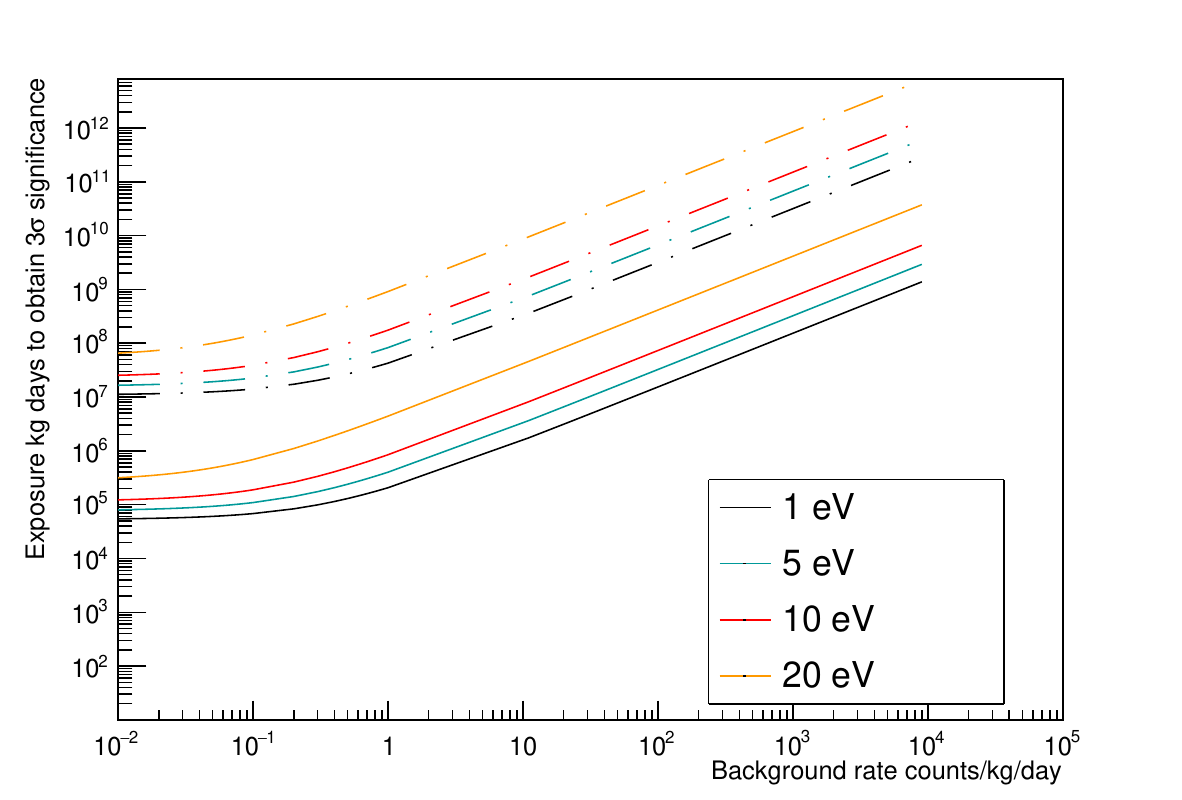}
    \caption{The exposure time required to reach a sensitivity of $3\sigma$ for various thresholds and background levels when the nearest or farthest CANSTOR baskets are removed for a pure tungsten target. The solid line gives the exposure to obtain sensitivity to the basket being removed that is closest to the detector. The dashed line gives the exposure for sensitivity to the farthest basket being removed.}
    \label{fig:PureTungstenSens}
\end{figure}

The exposure required to reach a sensitivity of $Z=3$ for a tungsten target for various background levels and various baskets being removed is shown in Figure~\ref{fig:PureTungstenSens}.

When a level of $Z=3$ is required, in the best-case scenario the measurement exposure is more than 150 kg years. Assuming a one-sided Gaussian fluctuation the probability of a $3\sigma$ fluctuation is $1.35\times10^{-3}$. As the false alarm rate can be found by dividing the measurement time by the fluctuation probability which results in a false alarm rate approximately once every 100,000 years for a 1~kg detector, the time associated with this false alarm rate is significantly longer than the duration of fuel storage within the container. Therefore, such a high level of sensitivity is not warranted when a 1~kg detector is used to monitor the spent fuel.

\begin{figure}
    \centering
    \includegraphics[width=\linewidth]{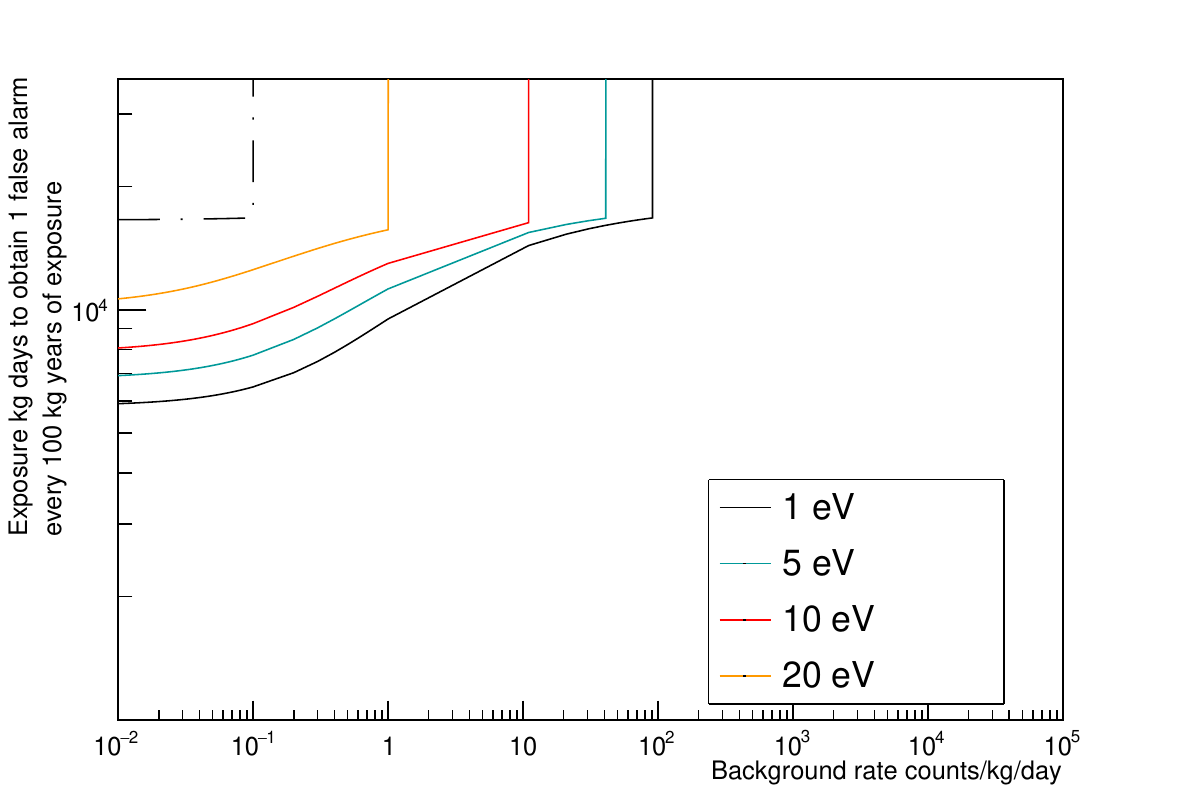}
    \caption{The measurement time required for a false alarm rate of once every 100 years for various background levels and thresholds when a pure tungsten target is used. The solid lines represent the scenario where the basket nearest to the detector is removed, and the dashed lines refer to the scenario where the basket farthest from the detector is removed.}
    \label{fig:falseAlarmRate}
\end{figure}

The measurement time required to reach an false alarm of once every 100 years of running for a 1~kg detector is shown in Figure~\ref{fig:falseAlarmRate}. For a single basket the best-case scenario where the closest basket is removed, a measurement can be made approximately once every 16 years for a 1 kg detector. If the measurement time is to be greatly shortened, a large detector is required. The measurement period for the removal of the farthest baskets is longer than 15,000 days, implying that only about two measurements can be taken every 100 years. In addition, the statistical uncertainty in these observations is so high that one out of these two will be a false alarm. For thresholds above 1~eV no curves are shown for the furthest basket being removed as the measurement time means only one measurement can be made in a 100 year period. As greatly enlarging the detector's size is impractical, numerous TES-based detectors will be required to monitor the entire container if sensitivity to removal of a single basket is required over an acceptable time-period. Note that a single basket corresponds to only approximately 0.5\% of the total fuel in the container. The need for a long exposure time to discriminate between a full container and one that is 99.5\% full is not unexpected given the low rate of neutrino events.

\subsection{Removal of an entire tube}
\begin{figure}
    \centering
    \includegraphics[width=\linewidth]{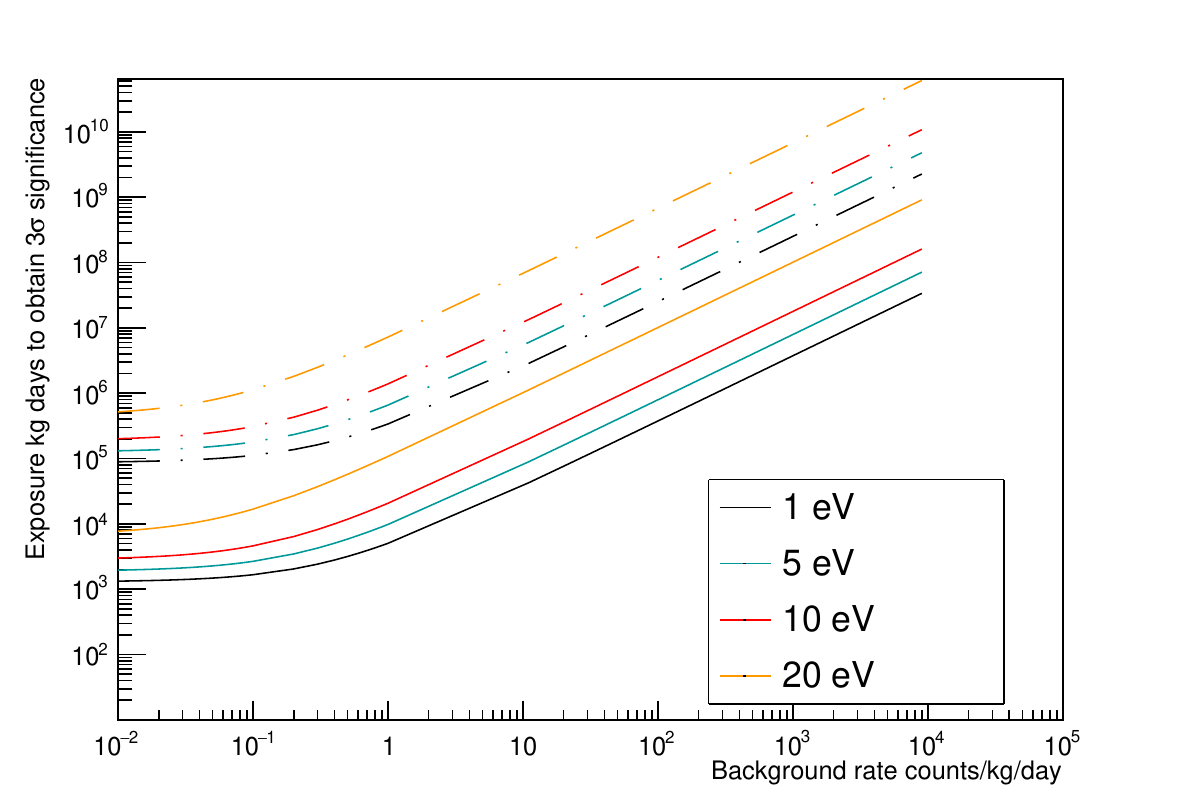}      
    \caption{The exposure required to reach $3\sigma$ sensitivity as a function of background level for various thresholds when various CANSTOR tubes are removed for a pure tungsten target. The solid lines give the exposure to obtain sensitivity to the tube being emptied that is closest to the detector. The dashed line gives the exposure for sensitivity to the farthest tube being emptied.}
    \label{fig:sensitivityEntireTube}
\end{figure}

The time required to reach a sensitivity of $Z=3$ when an entire tube is removed is shown in Figure~\ref{fig:sensitivityEntireTube}. For the tube being emptied that is closest to the detector an exposure of approximately 3.6 kg years is required to build up $3\sigma$ sensitivity if the background rate is 0.01 counts per day and the threshold of the detector is 1~eV.  However, the measurement exposure rises by up to 245 kg years when the tube farthest tube from the detector is emptied. 

\begin{figure}
    \centering
    \includegraphics[width=\linewidth]{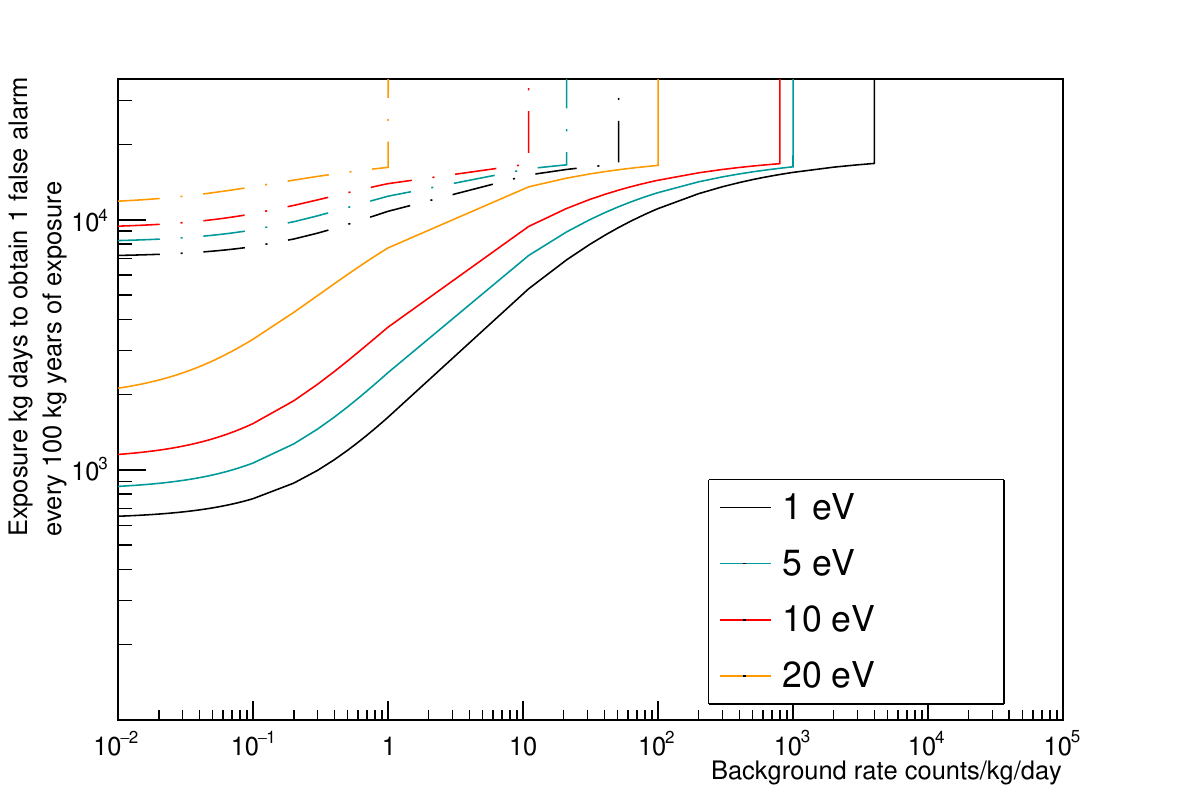}
    \caption{The measurement time required to result in a false alarm rate of once every 100 years for an empty tube in the container for various background levels and thresholds when a pure tungsten target is used. The solid lines represent the scenario where the tube nearest to the detector is empty and the dashed lines refer to the scenario where the farthest tube from the detector is empty.}
    \label{fig:falseAlarmRateEntireTube}
\end{figure}

The measurement time required to reach a false alarm rate of one per century kg of exposure for entire tubes being emptied is shown in Figure~\ref{fig:falseAlarmRateEntireTube}. For the tube being emptied that is closest to the detector the measurement period needs approximately 652 kg days with a 1~eV threshold and 0.01 background events per day. In contrast, when the tube farthest from the detector is emptied the measurement exposure is increased to 7,200 kg days.

\subsection{Using the TES detector to monitor a DSC}
\begin{figure}
    \centering
    \includegraphics[width=\linewidth]{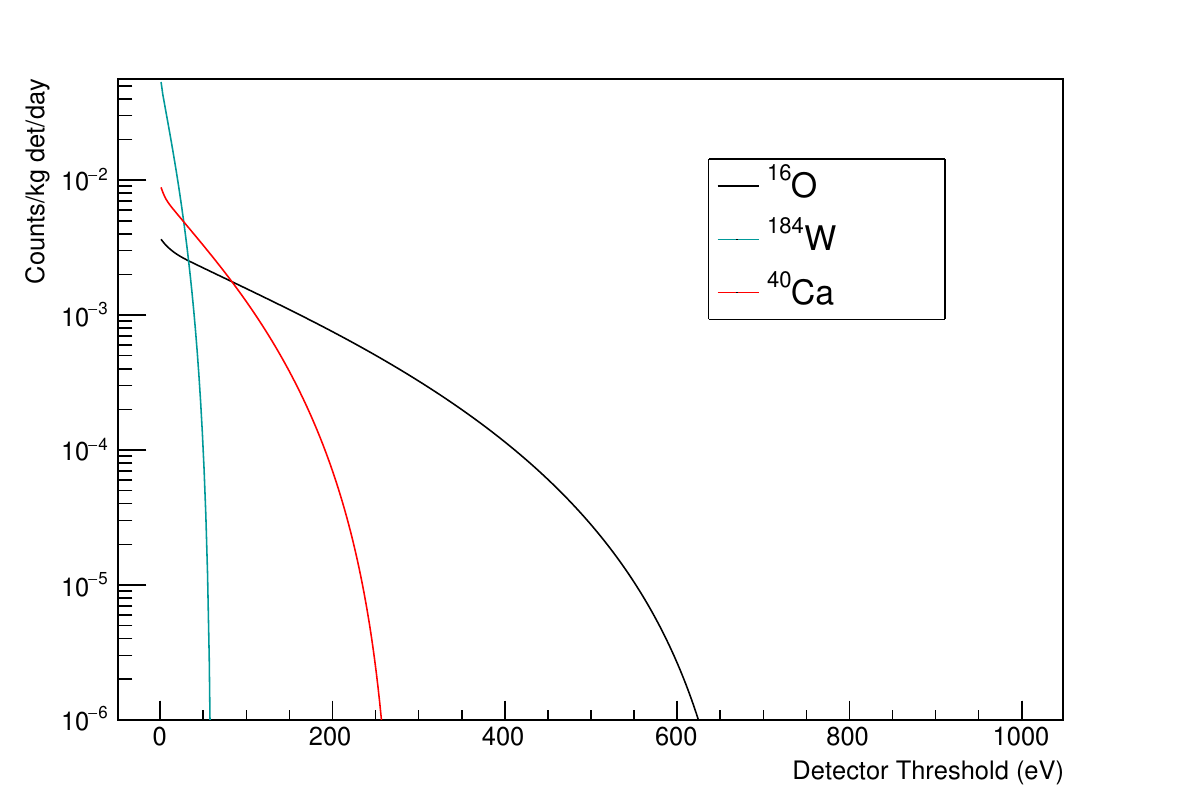}
    \caption{The rate of events for various materials for a sensor placed 2~m from a DSC full of 30-year-old spent fuel as a function of detector threshold.}
    \label{fig:DSCRateThreshold}
\end{figure}

Assuming the same 30-year-old fuel used in the CANSTOR studies is inside the DSC, the overall differential rate can be given by Eq.~\ref{eq:pointSourceRate}. The rate of events due to spent fuel in a full DSC as a function of detector threshold can be seen in Figure~\ref{fig:DSCRateThreshold}.

Modifying Eq.~\ref{eq:differenceInRate} such that $\mu_m = f\mu_F$, where $f$ is the fraction of fuel in the container, one obtains Eq.~\ref{eq:fractionFull}, which can be rearranged into Eq.~\ref{eq:runTimeFraction}.

\begin{equation}
    (\mu_F+b)t-(f\mu_F+b)t = Z\sqrt{(\mu_F+b)t}
    \label{eq:fractionFull}
\end{equation}

\begin{equation}
    t = \frac{Z^2(\mu_F+b)}{(1-f)^2\mu_F^2}
    \label{eq:runTimeFraction}
\end{equation}

\begin{figure*}
    \centering
    \includegraphics[width=\linewidth]{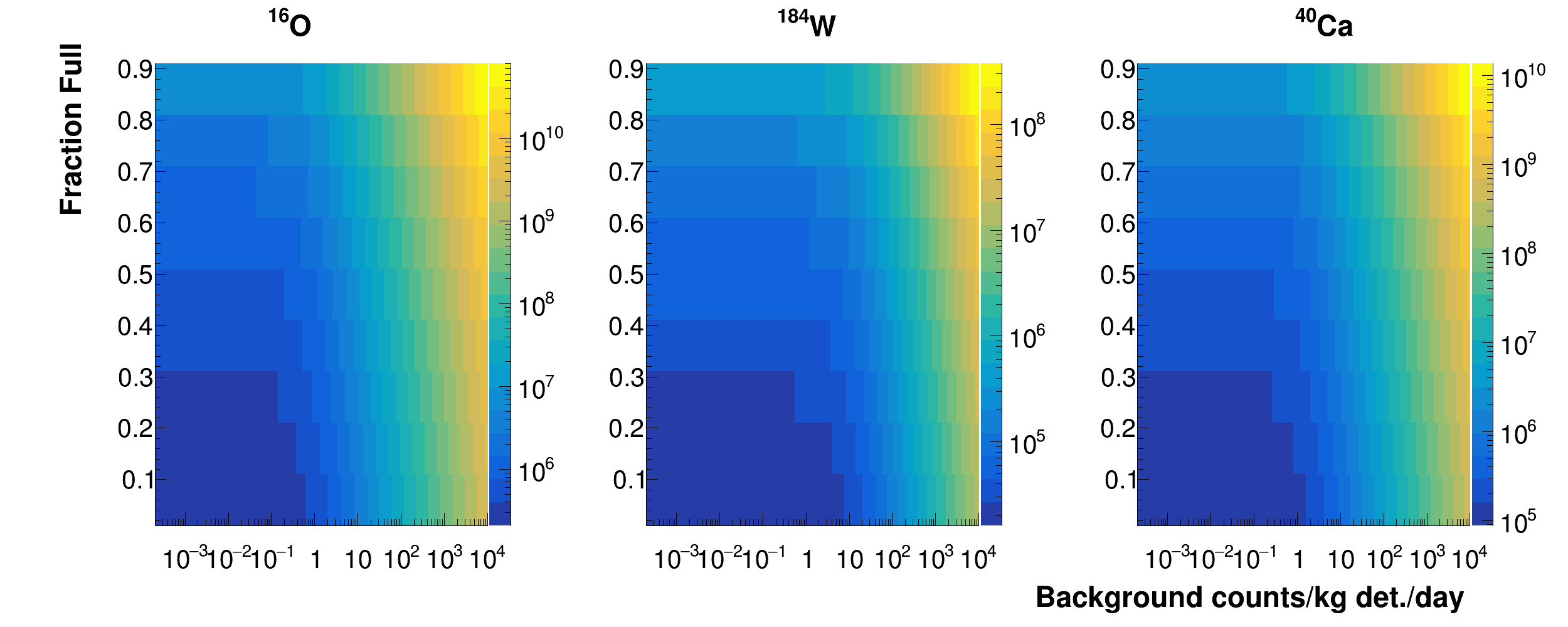}
    \caption{The exposure in kg days required to obtain a 95\% confidence that the detector is not full for various thresholds and target materials. The $x$ axis is the background rate ($\beta$) in counts per kg det. per day. The $y$ axis is the fraction of the DSC that is full. Each column corresponds to a different detector material. A 1~eV threshold is used for all materials.}
    \label{fig:DSCSensitivity}
\end{figure*}

Figure~\ref{fig:DSCSensitivity} shows the time to reach 95\% confidence that the container is not full for various thresholds and materials. The background rate $b$ in Eq.~\ref{eq:runTimeFraction} is given by
$b=\beta+\beta_\nu(\tau)$, where $\beta$ is the background rate as considered in the CANSTOR scenario and $\beta_\nu(\tau)$ is the rate of neutrino events due to the nearby reactors for a given threshold $\tau$ as listed in Table~\ref{tab:rateThresholds}.

Regardless of the target used, a significant exposure time is required to monitor an individual DSC. The target with the largest cross-section ($^{184}$W) and the lowest threshold requires almost 100,000 days of monitoring in order to determine whether the DSC is less than half full. Monitoring individual containers for such a large amount of time is impractical. Part of the reason for such a significant run time is the high rate of the neutrino events from the reactors. For a 20~eV threshold and a tungsten target the rate of neutrino events from the fuel is less than one per kg$^{-1}$ day$^{-1}$; however, the number of background events due to reactor neutrinos is 6.94 kg$^{-1}$ day$^{-1}$. Note that the recoils from the spent fuel are lower in energy than the recoils due to the reactor neutrinos; therefore, this background can  be mitigated using an upper energy cut. Table~\ref{tab:upperRecoilEnergy} shows the upper energy cut which would keep 99\% of the spent fuel recoils when the detector threshold is 1~eV. Applying this upper energy cut to the reactor backgrounds shown in Table~\ref{tab:rateThresholds} results in a reduction of rate seen in the third column of Table~\ref{tab:upperRecoilEnergy}, for a 1~eV threshold this corresponds to a reduction of the reactor background to 63\% of its pre-cut value. Even with this correction the rate due to the reactors is several orders of magnitude higher than the rate of neutrinos and the measurement time remains significantly higher than practically feasible. 

\begin{table}
    \centering
    \begin{tabular}{c|c|c}
         \textbf{Target} & \textbf{Energy Cut (eV)} & \thead{Reduction in \\ Reactor Rate} \\ \hline
         $^{16}$O & 484.1 & -0.26 \\ 
         $^{184}$W & 43.0 & -4.36\\
         $^{40}$Ca & 194.4 & -0.64\\
    \end{tabular}
    \caption{The recoil energy where 99\% of the interactions with energies above 1~eV occur and the reduction in the reactor background  neutrino background (in counts/kg det./day) if this cut is applied.}
    \label{tab:upperRecoilEnergy}
\end{table}

\section{Discussion}
The sensitivity studies in the previous section explored detectors on a $\sim$kg scale. Currently planned bolometers for neutrino detection range from a few grams to a kilogram scale with thresholds below 10~eV~\cite{Wagner_2021}. The threshold of the detector increases proportionally as $M^{2/3}$, where $M$ is the target size~\cite{Strauss_2017}. To compensate for this, as the target gets larger the operating temperature of the detector needs to be reduced to keep thermal noise at the same level and to maintain the same threshold~\cite{PhysRevD.85.013009}. Alternative options are to segment the detector into multiple smaller detectors~\cite{PhysRevD.85.013009}~\cite{PhysRevLett.55.25}. The segmentation of detectors could also enable improved background rejection~\cite{PhysRevLett.55.25}. Readout systems for X-ray TES based detection for are currently being developed with approximately ~3000 TES based detectors~\cite{Gottardi_2014} hence the segmentation of the detector will not add unreasonable complexity to the readout systems.

A further challenge to neutrino detection is the monitoring at the surface. The estimated background rate for an un-sheilded detector on the surface is 10,000 counts per day~\cite{PhysRevD.105.056002}. Significant effort will be required to evaluate and identify backgrounds and to design the detector to mitigate these. Future studies into fuel monitoring can use background measurements made by detectors with similar designs and locations such as NUCLEUS~\cite{Wagner_2021} and Ricochet~\cite{Ricochet:2021rjo}. These background calculations will also inform the design of the detector. An additional option would be to install the detector underground near the site, although this would significantly increase the construction complexity and overall detector cost, which may not be feasible around spent fuel containers.

\section{Conclusions}
The feasibility of using a TES based detector to monitor spent CANDU fuel was explored in two scenarios. The first of these was the monitoring of a CANSTOR container at the Gentilly-2 site. The detector was insensitive to a single basket being removed; when the basket closest to the detector is removed, a measurement time of 150 years with a 1 kg detector is required to obtain a signal with $3\sigma$ significance and more than $10^7$ days for the basket being removed that is farthest from the detector. It should be noted that a single basket corresponds to 0.5\% of the total fuel in a CANSTOR container, so a long exposure is required to distinguish such a small difference in rate.  The results are more promising when an entire tube of the CANSTOR container is emptied (which corresponds to ~5\% of the total fuel in the container). When the tube nearest to the detector is emptied for a 1 kg detector a total measurement time of 3.6 years is required in order to obtain $3\sigma$ significance from a full container. However, owing to geometric effects this measurement time is increased to 245 years when the tube farthest from the detector is emptied. In order to monitor a entire CANSTOR container, multiple detectors with sizes on the order of tens of kilograms located around the container to minimize geometric effects would be required.

The second scenario explored was the monitoring of an individual DSC located at the Pickering site. In this scenario there are additional backgrounds caused by the neutrinos produced by the nearby reactors; the rate of these backgrounds are much higher than the rate of events due to the fuel in the container. The combination of the high neutrino rate from the reactors as the relatively small amount of spent fuel in the DSC makes monitoring DSC using their neutrino emissions currently unfeasible. 

In both scenarios the exposure in kg days was explored. Detectors with sizes on the order of 10's of kilograms are required to monitor the spent fuel in a CANSTOR container within a reasonable time frame. Detectors of this size are not unfeasible with the development of a 1~kg scale detector planned in the near future~\cite{Wagner_2021}. As the target size limits the energy threshold of the detector, current detectors use multiple target arrays. The use of multiple arrays also aids in discriminating background events. Furthermore, this modularity means that a shorter monitoring time can be achieved by simply adding more targets provided there is room in the cryostat.
\section{Acknowledgements}
We gratefully acknowledge the insightful exchanges with Bhaskar Sur and his suggestion to explore the use of TES-based detectors for neutrino detection and nuclear recoil measurements. We also thank Oleg Kamaev for his careful review of this paper.

This work is funded by the Federal Nuclear Science \& Technology fund through Atomic Energy Canada Limited (AECL). We appreciate all the help and support provided by AECL.


\begin{thebibliography}{33}%
\makeatletter
\providecommand \@ifxundefined [1]{%
 \@ifx{#1\undefined}
}%
\providecommand \@ifnum [1]{%
 \ifnum #1\expandafter \@firstoftwo
 \else \expandafter \@secondoftwo
 \fi
}%
\providecommand \@ifx [1]{%
 \ifx #1\expandafter \@firstoftwo
 \else \expandafter \@secondoftwo
 \fi
}%
\providecommand \natexlab [1]{#1}%
\providecommand \enquote  [1]{``#1''}%
\providecommand \bibnamefont  [1]{#1}%
\providecommand \bibfnamefont [1]{#1}%
\providecommand \citenamefont [1]{#1}%
\providecommand \href@noop [0]{\@secondoftwo}%
\providecommand \href [0]{\begingroup \@sanitize@url \@href}%
\providecommand \@href[1]{\@@startlink{#1}\@@href}%
\providecommand \@@href[1]{\endgroup#1\@@endlink}%
\providecommand \@sanitize@url [0]{\catcode `\\12\catcode `\$12\catcode
  `\&12\catcode `\#12\catcode `\^12\catcode `\_12\catcode `\%12\relax}%
\providecommand \@@startlink[1]{}%
\providecommand \@@endlink[0]{}%
\providecommand \url  [0]{\begingroup\@sanitize@url \@url }%
\providecommand \@url [1]{\endgroup\@href {#1}{\urlprefix }}%
\providecommand \urlprefix  [0]{URL }%
\providecommand \Eprint [0]{\href }%
\providecommand \doibase [0]{https://doi.org/}%
\providecommand \selectlanguage [0]{\@gobble}%
\providecommand \bibinfo  [0]{\@secondoftwo}%
\providecommand \bibfield  [0]{\@secondoftwo}%
\providecommand \translation [1]{[#1]}%
\providecommand \BibitemOpen [0]{}%
\providecommand \bibitemStop [0]{}%
\providecommand \bibitemNoStop [0]{.\EOS\space}%
\providecommand \EOS [0]{\spacefactor3000\relax}%
\providecommand \BibitemShut  [1]{\csname bibitem#1\endcsname}%
\let\auto@bib@innerbib\@empty
\bibitem [{\citenamefont {O.Akindele}\ \emph {et~al.}(2021)\citenamefont
  {O.Akindele}, \citenamefont {N.Bowden}, \citenamefont {R.Carr}, \citenamefont
  {A.Conant}, \citenamefont {M.Diwan}, \citenamefont {A.Erickson},
  \citenamefont {M.Foxe}, \citenamefont {B.L.Goldblum}, \citenamefont
  {P.Huber}, \citenamefont {I.Jovanovic} \emph {et~al.}}]{2021arXiv211212593A}%
  \BibitemOpen
  \bibfield  {author} {\bibinfo {author} {\bibnamefont {O.Akindele}}, \bibinfo
  {author} {\bibnamefont {N.Bowden}}, \bibinfo {author} {\bibnamefont
  {R.Carr}}, \bibinfo {author} {\bibnamefont {A.Conant}}, \bibinfo {author}
  {\bibnamefont {M.Diwan}}, \bibinfo {author} {\bibnamefont {A.Erickson}},
  \bibinfo {author} {\bibnamefont {M.Foxe}}, \bibinfo {author} {\bibnamefont
  {B.L.Goldblum}}, \bibinfo {author} {\bibnamefont {P.Huber}}, \bibinfo
  {author} {\bibnamefont {I.Jovanovic}}, \emph {et~al.},\ }\href
  {https://doi.org/10.2172/1826602} {\emph {\bibinfo {title} {Nu Tools:
  Exploring Practical Roles for Neutrinos in Nuclear Energy and Security}}},\
  \bibinfo {type} {Tech. Rep.}\ (\bibinfo {year} {2021})\BibitemShut {NoStop}%
\bibitem [{\citenamefont {Raesfeld}\ and\ \citenamefont
  {P.Huber}(2022)}]{PhysRevD.105.056002}%
  \BibitemOpen
  \bibfield  {author} {\bibinfo {author} {\bibfnamefont {C.}~\bibnamefont
  {Raesfeld}}\ and\ \bibinfo {author} {\bibnamefont {P.Huber}},\ }\bibfield
  {title} {\bibinfo {title} {Use of {CEvNS} to monitor spent nuclear fuel},\
  }\href {https://doi.org/10.1103/PhysRevD.105.056002} {\bibfield  {journal}
  {\bibinfo  {journal} {Phys. Rev. D}\ }\textbf {\bibinfo {volume} {105}},\
  \bibinfo {pages} {056002} (\bibinfo {year} {2022})}\BibitemShut {NoStop}%
\bibitem [{\citenamefont {V.Brdar}\ \emph {et~al.}(2017)\citenamefont
  {V.Brdar}, \citenamefont {P.Huber},\ and\ \citenamefont
  {J.Kopp}}]{PhysRevApplied.8.054050}%
  \BibitemOpen
  \bibfield  {author} {\bibinfo {author} {\bibnamefont {V.Brdar}}, \bibinfo
  {author} {\bibnamefont {P.Huber}},\ and\ \bibinfo {author} {\bibnamefont
  {J.Kopp}},\ }\bibfield  {title} {\bibinfo {title} {Antineutrino monitoring of
  spent nuclear fuel},\ }\href
  {https://doi.org/10.1103/PhysRevApplied.8.054050} {\bibfield  {journal}
  {\bibinfo  {journal} {Phys. Rev. Appl.}\ }\textbf {\bibinfo {volume} {8}},\
  \bibinfo {pages} {054050} (\bibinfo {year} {2017})}\BibitemShut {NoStop}%
\bibitem [{\citenamefont {Irwin}\ and\ \citenamefont
  {Hilton}(2005)}]{Irwin2005}%
  \BibitemOpen
  \bibfield  {author} {\bibinfo {author} {\bibfnamefont {K.}~\bibnamefont
  {Irwin}}\ and\ \bibinfo {author} {\bibfnamefont {G.}~\bibnamefont {Hilton}},\
  }\bibinfo {title} {Transition-edge sensors},\ in\ \href
  {https://doi.org/10.1007/10933596_3} {\emph {\bibinfo {booktitle} {Cryogenic
  Particle Detection}}}\ (\bibinfo  {publisher} {Springer Berlin Heidelberg},\
  \bibinfo {address} {Berlin, Heidelberg},\ \bibinfo {year} {2005})\ pp.\
  \bibinfo {pages} {63--150}\BibitemShut {NoStop}%
\bibitem [{\citenamefont {K.Scholberg}(2018)}]{Scholberg:2018vwg}%
  \BibitemOpen
  \bibfield  {author} {\bibinfo {author} {\bibnamefont {K.Scholberg}} (\bibinfo
  {collaboration} {COHERENT}),\ }\bibfield  {title} {\bibinfo {title}
  {{Observation of Coherent Elastic Neutrino-Nucleus Scattering by COHERENT}},\
  }\href {https://doi.org/10.22323/1.295.0020} {\bibfield  {journal} {\bibinfo
  {journal} {PoS}\ }\textbf {\bibinfo {volume} {NuFact2017}},\ \bibinfo {pages}
  {020} (\bibinfo {year} {2018})},\ \Eprint {https://arxiv.org/abs/1801.05546}
  {arXiv:1801.05546 [hep-ex]} \BibitemShut {NoStop}%
\bibitem [{\citenamefont {Akimov}\ \emph {et~al.}(2017)\citenamefont {Akimov},
  \citenamefont {Albert}, \citenamefont {An}, \citenamefont {Awe},
  \citenamefont {Barbeau}, \citenamefont {Becker}, \citenamefont {Belov},
  \citenamefont {Brown}, \citenamefont {Bolozdynya}, \citenamefont
  {Cabrera-Palmer} \emph {et~al.}}]{doi:10.1126/science.aao0990}%
  \BibitemOpen
  \bibfield  {author} {\bibinfo {author} {\bibfnamefont {D.}~\bibnamefont
  {Akimov}}, \bibinfo {author} {\bibfnamefont {J.~B.}\ \bibnamefont {Albert}},
  \bibinfo {author} {\bibfnamefont {P.}~\bibnamefont {An}}, \bibinfo {author}
  {\bibfnamefont {C.}~\bibnamefont {Awe}}, \bibinfo {author} {\bibfnamefont
  {P.~S.}\ \bibnamefont {Barbeau}}, \bibinfo {author} {\bibfnamefont
  {B.}~\bibnamefont {Becker}}, \bibinfo {author} {\bibfnamefont
  {V.}~\bibnamefont {Belov}}, \bibinfo {author} {\bibfnamefont
  {A.}~\bibnamefont {Brown}}, \bibinfo {author} {\bibfnamefont
  {A.}~\bibnamefont {Bolozdynya}}, \bibinfo {author} {\bibfnamefont
  {B.}~\bibnamefont {Cabrera-Palmer}}, \emph {et~al.},\ }\bibfield  {title}
  {\bibinfo {title} {Observation of coherent elastic neutrino-nucleus
  scattering},\ }\href {https://doi.org/10.1126/science.aao0990} {\bibfield
  {journal} {\bibinfo  {journal} {Science}\ }\textbf {\bibinfo {volume}
  {357}},\ \bibinfo {pages} {1123} (\bibinfo {year} {2017})},\ \Eprint
  {https://arxiv.org/abs/https://www.science.org/doi/pdf/10.1126/science.aao0990}
  {https://www.science.org/doi/pdf/10.1126/science.aao0990} \BibitemShut
  {NoStop}%
\bibitem [{\citenamefont {Akimov}\ \emph {et~al.}(2021)\citenamefont {Akimov},
  \citenamefont {Albert}, \citenamefont {An}, \citenamefont {Awe},
  \citenamefont {Barbeau}, \citenamefont {Becker}, \citenamefont {Belov},
  \citenamefont {Bernardi}, \citenamefont {Blackston}, \citenamefont {Blokland}
  \emph {et~al.}}]{PhysRevLett.126.012002}%
  \BibitemOpen
  \bibfield  {author} {\bibinfo {author} {\bibfnamefont {D.}~\bibnamefont
  {Akimov}}, \bibinfo {author} {\bibfnamefont {J.~B.}\ \bibnamefont {Albert}},
  \bibinfo {author} {\bibfnamefont {P.}~\bibnamefont {An}}, \bibinfo {author}
  {\bibfnamefont {C.}~\bibnamefont {Awe}}, \bibinfo {author} {\bibfnamefont
  {P.~S.}\ \bibnamefont {Barbeau}}, \bibinfo {author} {\bibfnamefont
  {B.}~\bibnamefont {Becker}}, \bibinfo {author} {\bibfnamefont
  {V.}~\bibnamefont {Belov}}, \bibinfo {author} {\bibfnamefont
  {I.}~\bibnamefont {Bernardi}}, \bibinfo {author} {\bibfnamefont {M.~A.}\
  \bibnamefont {Blackston}}, \bibinfo {author} {\bibfnamefont {L.}~\bibnamefont
  {Blokland}}, \emph {et~al.} (\bibinfo {collaboration} {COHERENT
  Collaboration}),\ }\bibfield  {title} {\bibinfo {title} {First measurement of
  coherent elastic neutrino-nucleus scattering on argon},\ }\href
  {https://doi.org/10.1103/PhysRevLett.126.012002} {\bibfield  {journal}
  {\bibinfo  {journal} {Phys. Rev. Lett.}\ }\textbf {\bibinfo {volume} {126}},\
  \bibinfo {pages} {012002} (\bibinfo {year} {2021})}\BibitemShut {NoStop}%
\bibitem [{\citenamefont {M.Tayal}\ and\ \citenamefont
  {M.Gacesa}(2017)}]{CANDUFuel}%
  \BibitemOpen
  \bibfield  {author} {\bibinfo {author} {\bibnamefont {M.Tayal}}\ and\
  \bibinfo {author} {\bibnamefont {M.Gacesa}},\ }\bibfield  {title} {\bibinfo
  {title} {Fuel},\ }in\ \href@noop {} {\emph {\bibinfo {booktitle} {The
  Essential CANDU}}}\ (\bibinfo  {publisher} {UNENE},\ \bibinfo {year} {2017})\
  Chap.~\bibinfo {chapter} {17}\BibitemShut {NoStop}%
\bibitem [{\citenamefont {V.Wagner}\ \emph {et~al.}(2021)\citenamefont
  {V.Wagner}, \citenamefont {G.Angloher}, \citenamefont {A.Bento},
  \citenamefont {L.Canonica}, \citenamefont {F.Cappella}, \citenamefont
  {L.Cardani}, \citenamefont {N.Casali}, \citenamefont {R.Cerulli},
  \citenamefont {I.Colantoni}, \citenamefont {A.Cruciani} \emph
  {et~al.}}]{Wagner_2021}%
  \BibitemOpen
  \bibfield  {author} {\bibinfo {author} {\bibnamefont {V.Wagner}}, \bibinfo
  {author} {\bibnamefont {G.Angloher}}, \bibinfo {author} {\bibnamefont
  {A.Bento}}, \bibinfo {author} {\bibnamefont {L.Canonica}}, \bibinfo {author}
  {\bibnamefont {F.Cappella}}, \bibinfo {author} {\bibnamefont {L.Cardani}},
  \bibinfo {author} {\bibnamefont {N.Casali}}, \bibinfo {author} {\bibnamefont
  {R.Cerulli}}, \bibinfo {author} {\bibnamefont {I.Colantoni}}, \bibinfo
  {author} {\bibnamefont {A.Cruciani}}, \emph {et~al.},\ }\bibfield  {title}
  {\bibinfo {title} {Exploring {CE$\nu$NS} of reactor neutrinos with the
  nucleus experiment},\ }\href
  {https://doi.org/10.1088/1742-6596/2156/1/012118} {\bibfield  {journal}
  {\bibinfo  {journal} {J. Phys. Conf. Ser.}\ }\textbf {\bibinfo {volume}
  {2156}},\ \bibinfo {pages} {012118} (\bibinfo {year} {2021})}\BibitemShut
  {NoStop}%
\bibitem [{\citenamefont {Macici}(1997)}]{CANSTORInfo}%
  \BibitemOpen
  \bibfield  {author} {\bibinfo {author} {\bibfnamefont {N.}~\bibnamefont
  {Macici}},\ }\href
  {https://inis.iaea.org/collection/NCLCollectionStore/_Public/31/006/31006110.pdf}
  {\emph {\bibinfo {title} {Spent Fuel Dry Storage Experience At {Gentilly-2
  NGS}}}},\ \bibinfo {type} {Tech. Rep.}\ \bibinfo {number} {CA0000091}\
  (\bibinfo  {institution} {Hydro-Qu\`ebec},\ \bibinfo {year}
  {1997})\BibitemShut {NoStop}%
\bibitem [{\citenamefont {Pattantyus}(1998)}]{DSCInfo}%
  \BibitemOpen
  \bibfield  {author} {\bibinfo {author} {\bibfnamefont {P.}~\bibnamefont
  {Pattantyus}},\ }\href
  {https://inis.iaea.org/collection/NCLCollectionStore/_Public/29/026/29026634.pdf}
  {\emph {\bibinfo {title} {{Spent Fuel Management In Canada}}}},\ \bibinfo
  {type} {Tech. Rep.}\ \bibinfo {number} {XA9846847}\ (\bibinfo  {institution}
  {AECL},\ \bibinfo {year} {1998})\BibitemShut {NoStop}%
\bibitem [{\citenamefont {O'Hare}(2016)}]{PhysRevD.94.063527}%
  \BibitemOpen
  \bibfield  {author} {\bibinfo {author} {\bibfnamefont {C.}~\bibnamefont
  {O'Hare}},\ }\bibfield  {title} {\bibinfo {title} {Dark matter astrophysical
  uncertainties and the neutrino floor},\ }\href
  {https://doi.org/10.1103/PhysRevD.94.063527} {\bibfield  {journal} {\bibinfo
  {journal} {Phys. Rev. D}\ }\textbf {\bibinfo {volume} {94}},\ \bibinfo
  {pages} {063527} (\bibinfo {year} {2016})}\BibitemShut {NoStop}%
\bibitem [{\citenamefont {McCabe}(2010)}]{PhysRevD.82.023530}%
  \BibitemOpen
  \bibfield  {author} {\bibinfo {author} {\bibfnamefont {C.}~\bibnamefont
  {McCabe}},\ }\bibfield  {title} {\bibinfo {title} {Astrophysical
  uncertainties of dark matter direct detection experiments},\ }\href
  {https://doi.org/10.1103/PhysRevD.82.023530} {\bibfield  {journal} {\bibinfo
  {journal} {Phys. Rev. D}\ }\textbf {\bibinfo {volume} {82}},\ \bibinfo
  {pages} {023530} (\bibinfo {year} {2010})}\BibitemShut {NoStop}%
\bibitem [{\citenamefont {Pyle}\ \emph {et~al.}(2015)\citenamefont {Pyle},
  \citenamefont {Figueroa-Feliciano},\ and\ \citenamefont
  {Sadoulet}}]{pyle2015optimized}%
  \BibitemOpen
  \bibfield  {author} {\bibinfo {author} {\bibfnamefont {M.}~\bibnamefont
  {Pyle}}, \bibinfo {author} {\bibfnamefont {E.}~\bibnamefont
  {Figueroa-Feliciano}},\ and\ \bibinfo {author} {\bibfnamefont
  {B.}~\bibnamefont {Sadoulet}},\ }\href@noop {} {\bibinfo {title} {Optimized
  designs for very low temperature massive calorimeters}} (\bibinfo {year}
  {2015}),\ \Eprint {https://arxiv.org/abs/1503.01200} {arXiv:1503.01200
  [astro-ph.IM]} \BibitemShut {NoStop}%
\bibitem [{\citenamefont {Formaggio}\ \emph {et~al.}(2012)\citenamefont
  {Formaggio}, \citenamefont {Figueroa-Feliciano},\ and\ \citenamefont
  {Anderson}}]{PhysRevD.85.013009}%
  \BibitemOpen
  \bibfield  {author} {\bibinfo {author} {\bibfnamefont {J.}~\bibnamefont
  {Formaggio}}, \bibinfo {author} {\bibfnamefont {E.}~\bibnamefont
  {Figueroa-Feliciano}},\ and\ \bibinfo {author} {\bibfnamefont {A.~J.}\
  \bibnamefont {Anderson}},\ }\bibfield  {title} {\bibinfo {title} {Sterile
  neutrinos, coherent scattering, and oscillometry measurements with
  low-temperature bolometers},\ }\href
  {https://doi.org/10.1103/PhysRevD.85.013009} {\bibfield  {journal} {\bibinfo
  {journal} {Phys. Rev. D}\ }\textbf {\bibinfo {volume} {85}},\ \bibinfo
  {pages} {013009} (\bibinfo {year} {2012})}\BibitemShut {NoStop}%
\bibitem [{\citenamefont {{Strauss}}\ \emph {et~al.}(2017)\citenamefont
  {{Strauss}}, \citenamefont {{Rothe}}, \citenamefont {{Angloher}},
  \citenamefont {{Bento}}, \citenamefont {{G{\"u}tlein}}, \citenamefont
  {{Hauff}}, \citenamefont {{Kluck}}, \citenamefont {{Mancuso}}, \citenamefont
  {{Oberauer}}, \citenamefont {{Petricca}} \emph
  {et~al.}}]{2017EPJC...77..506S}%
  \BibitemOpen
  \bibfield  {author} {\bibinfo {author} {\bibfnamefont {R.}~\bibnamefont
  {{Strauss}}}, \bibinfo {author} {\bibfnamefont {J.}~\bibnamefont {{Rothe}}},
  \bibinfo {author} {\bibfnamefont {G.}~\bibnamefont {{Angloher}}}, \bibinfo
  {author} {\bibfnamefont {A.}~\bibnamefont {{Bento}}}, \bibinfo {author}
  {\bibfnamefont {A.}~\bibnamefont {{G{\"u}tlein}}}, \bibinfo {author}
  {\bibfnamefont {D.}~\bibnamefont {{Hauff}}}, \bibinfo {author} {\bibfnamefont
  {H.}~\bibnamefont {{Kluck}}}, \bibinfo {author} {\bibfnamefont
  {M.}~\bibnamefont {{Mancuso}}}, \bibinfo {author} {\bibfnamefont
  {L.}~\bibnamefont {{Oberauer}}}, \bibinfo {author} {\bibfnamefont
  {F.}~\bibnamefont {{Petricca}}}, \emph {et~al.},\ }\bibfield  {title}
  {\bibinfo {title} {{The {\ensuremath{\nu}}-cleus experiment: a gram-scale
  fiducial-volume cryogenic detector for the first detection of coherent
  neutrino-nucleus scattering}},\ }\href
  {https://doi.org/10.1140/epjc/s10052-017-5068-2} {\bibfield  {journal}
  {\bibinfo  {journal} {European Physical Journal C}\ }\textbf {\bibinfo
  {volume} {77}},\ \bibinfo {eid} {506} (\bibinfo {year} {2017})},\ \Eprint
  {https://arxiv.org/abs/1704.04320} {arXiv:1704.04320 [physics.ins-det]}
  \BibitemShut {NoStop}%
\bibitem [{\citenamefont {Heckman}\ and\ \citenamefont
  {Edward}(2020)}]{NWMOSpentFuelComp}%
  \BibitemOpen
  \bibfield  {author} {\bibinfo {author} {\bibfnamefont {K.}~\bibnamefont
  {Heckman}}\ and\ \bibinfo {author} {\bibfnamefont {J.}~\bibnamefont
  {Edward}},\ }\href
  {https://www.nwmo.ca/~/media/Site/Reports/2020/08/28/19/52/NWMOTR202005.ashx?la=en}
  {\emph {\bibinfo {title} {Radionuclide Inventory for Reference {CANDU} Fuel
  Bundles}}},\ \bibinfo {type} {Tech. Rep.}\ \bibinfo {number}
  {NWMO-TR-2020-05}\ (\bibinfo  {institution} {Nuclear Waste Management
  Organization},\ \bibinfo {year} {2020})\BibitemShut {NoStop}%
\bibitem [{\citenamefont {B\'e}\ \emph
  {et~al.}(2004{\natexlab{a}})\citenamefont {B\'e}, \citenamefont {Chist\'e},
  \citenamefont {Dulieu}, \citenamefont {Browne}, \citenamefont {Chechev},
  \citenamefont {Kuzmenko}, \citenamefont {Helmer}, \citenamefont {Nichols},
  \citenamefont {Sch\"onfeld},\ and\ \citenamefont {Dersch}}]{TabRad_v1}%
  \BibitemOpen
  \bibfield  {author} {\bibinfo {author} {\bibfnamefont {M.-M.}\ \bibnamefont
  {B\'e}}, \bibinfo {author} {\bibfnamefont {V.}~\bibnamefont {Chist\'e}},
  \bibinfo {author} {\bibfnamefont {C.}~\bibnamefont {Dulieu}}, \bibinfo
  {author} {\bibfnamefont {E.}~\bibnamefont {Browne}}, \bibinfo {author}
  {\bibfnamefont {V.}~\bibnamefont {Chechev}}, \bibinfo {author} {\bibfnamefont
  {N.}~\bibnamefont {Kuzmenko}}, \bibinfo {author} {\bibfnamefont
  {R.}~\bibnamefont {Helmer}}, \bibinfo {author} {\bibfnamefont
  {A.}~\bibnamefont {Nichols}}, \bibinfo {author} {\bibfnamefont
  {E.}~\bibnamefont {Sch\"onfeld}},\ and\ \bibinfo {author} {\bibfnamefont
  {R.}~\bibnamefont {Dersch}},\ }\href
  {http://www.bipm.org/utils/common/pdf/monographieRI/Monographie_BIPM-5_Tables_Vol1.pdf}
  {\emph {\bibinfo {title} {Table of Radionuclides}}},\ \bibinfo {series}
  {Monographie BIPM-5}, Vol.~\bibinfo {volume} {1}\ (\bibinfo  {publisher}
  {Bureau International des Poids et Mesures},\ \bibinfo {address} {Pavillon de
  Breteuil, F-92310 S\`evres, France},\ \bibinfo {year} {2004})\BibitemShut
  {NoStop}%
\bibitem [{\citenamefont {B\'e}\ \emph
  {et~al.}(2004{\natexlab{b}})\citenamefont {B\'e}, \citenamefont {Chist\'e},
  \citenamefont {Dulieu}, \citenamefont {Browne}, \citenamefont {Chechev},
  \citenamefont {Kuzmenko}, \citenamefont {Helmer}, \citenamefont {Nichols},
  \citenamefont {Sch\"onfeld},\ and\ \citenamefont {Dersch}}]{TabRad_v2}%
  \BibitemOpen
  \bibfield  {author} {\bibinfo {author} {\bibfnamefont {M.-M.}\ \bibnamefont
  {B\'e}}, \bibinfo {author} {\bibfnamefont {V.}~\bibnamefont {Chist\'e}},
  \bibinfo {author} {\bibfnamefont {C.}~\bibnamefont {Dulieu}}, \bibinfo
  {author} {\bibfnamefont {E.}~\bibnamefont {Browne}}, \bibinfo {author}
  {\bibfnamefont {V.}~\bibnamefont {Chechev}}, \bibinfo {author} {\bibfnamefont
  {N.}~\bibnamefont {Kuzmenko}}, \bibinfo {author} {\bibfnamefont
  {R.}~\bibnamefont {Helmer}}, \bibinfo {author} {\bibfnamefont
  {A.}~\bibnamefont {Nichols}}, \bibinfo {author} {\bibfnamefont
  {E.}~\bibnamefont {Sch\"onfeld}},\ and\ \bibinfo {author} {\bibfnamefont
  {R.}~\bibnamefont {Dersch}},\ }\href
  {http://www.bipm.org/utils/common/pdf/monographieRI/Monographie_BIPM-5_Tables_Vol2.pdf}
  {\emph {\bibinfo {title} {Table of Radionuclides}}},\ \bibinfo {series}
  {Monographie BIPM-5}, Vol.~\bibinfo {volume} {2}\ (\bibinfo  {publisher}
  {Bureau International des Poids et Mesures},\ \bibinfo {address} {Pavillon de
  Breteuil, F-92310 S\`evres, France},\ \bibinfo {year} {2004})\BibitemShut
  {NoStop}%
\bibitem [{\citenamefont {B\'e}\ \emph {et~al.}(2006)\citenamefont {B\'e},
  \citenamefont {Chist\'e}, \citenamefont {Dulieu}, \citenamefont {Browne},
  \citenamefont {Baglin}, \citenamefont {Chechev}, \citenamefont {Kuzmenko},
  \citenamefont {Helmer}, \citenamefont {Kondev}, \citenamefont {MacMahon},\
  and\ \citenamefont {Lee}}]{TabRad_v3}%
  \BibitemOpen
  \bibfield  {author} {\bibinfo {author} {\bibfnamefont {M.-M.}\ \bibnamefont
  {B\'e}}, \bibinfo {author} {\bibfnamefont {V.}~\bibnamefont {Chist\'e}},
  \bibinfo {author} {\bibfnamefont {C.}~\bibnamefont {Dulieu}}, \bibinfo
  {author} {\bibfnamefont {E.}~\bibnamefont {Browne}}, \bibinfo {author}
  {\bibfnamefont {C.}~\bibnamefont {Baglin}}, \bibinfo {author} {\bibfnamefont
  {V.}~\bibnamefont {Chechev}}, \bibinfo {author} {\bibfnamefont
  {N.}~\bibnamefont {Kuzmenko}}, \bibinfo {author} {\bibfnamefont
  {R.}~\bibnamefont {Helmer}}, \bibinfo {author} {\bibfnamefont
  {F.}~\bibnamefont {Kondev}}, \bibinfo {author} {\bibfnamefont
  {D.}~\bibnamefont {MacMahon}},\ and\ \bibinfo {author} {\bibfnamefont
  {K.}~\bibnamefont {Lee}},\ }\href
  {http://www.bipm.org/utils/common/pdf/monographieRI/Monographie_BIPM-5_Tables_Vol3.pdf}
  {\emph {\bibinfo {title} {Table of Radionuclides}}},\ \bibinfo {series}
  {Monographie BIPM-5}, Vol.~\bibinfo {volume} {3}\ (\bibinfo  {publisher}
  {Bureau International des Poids et Mesures},\ \bibinfo {address} {Pavillon de
  Breteuil, F-92310 S\`evres, France},\ \bibinfo {year} {2006})\BibitemShut
  {NoStop}%
\bibitem [{\citenamefont {B\'e}\ \emph {et~al.}(2008)\citenamefont {B\'e},
  \citenamefont {Chist\'e}, \citenamefont {Dulieu}, \citenamefont {Browne},
  \citenamefont {Chechev}, \citenamefont {Kuzmenko}, \citenamefont {Kondev},
  \citenamefont {Luca}, \citenamefont {Gal\'an}, \citenamefont {Pearce},\ and\
  \citenamefont {Huang}}]{TabRad_v4}%
  \BibitemOpen
  \bibfield  {author} {\bibinfo {author} {\bibfnamefont {M.-M.}\ \bibnamefont
  {B\'e}}, \bibinfo {author} {\bibfnamefont {V.}~\bibnamefont {Chist\'e}},
  \bibinfo {author} {\bibfnamefont {C.}~\bibnamefont {Dulieu}}, \bibinfo
  {author} {\bibfnamefont {E.}~\bibnamefont {Browne}}, \bibinfo {author}
  {\bibfnamefont {V.}~\bibnamefont {Chechev}}, \bibinfo {author} {\bibfnamefont
  {N.}~\bibnamefont {Kuzmenko}}, \bibinfo {author} {\bibfnamefont
  {F.}~\bibnamefont {Kondev}}, \bibinfo {author} {\bibfnamefont
  {A.}~\bibnamefont {Luca}}, \bibinfo {author} {\bibfnamefont {M.}~\bibnamefont
  {Gal\'an}}, \bibinfo {author} {\bibfnamefont {A.}~\bibnamefont {Pearce}},\
  and\ \bibinfo {author} {\bibfnamefont {X.}~\bibnamefont {Huang}},\ }\href
  {http://www.bipm.org/utils/common/pdf/monographieRI/Monographie_BIPM-5_Tables_Vol4.pdf}
  {\emph {\bibinfo {title} {Table of Radionuclides}}},\ \bibinfo {series}
  {Monographie BIPM-5}, Vol.~\bibinfo {volume} {4}\ (\bibinfo  {publisher}
  {Bureau International des Poids et Mesures},\ \bibinfo {address} {Pavillon de
  Breteuil, F-92310 S\`evres, France},\ \bibinfo {year} {2008})\BibitemShut
  {NoStop}%
\bibitem [{\citenamefont {B\'e}\ \emph {et~al.}(2010)\citenamefont {B\'e},
  \citenamefont {Chist\'e}, \citenamefont {Dulieu}, \citenamefont {Mougeot},
  \citenamefont {Browne}, \citenamefont {Chechev}, \citenamefont {Kuzmenko},
  \citenamefont {Kondev}, \citenamefont {Luca}, \citenamefont {Gal\'an},
  \citenamefont {Nichols}, \citenamefont {Arinc},\ and\ \citenamefont
  {Huang}}]{TabRad_v5}%
  \BibitemOpen
  \bibfield  {author} {\bibinfo {author} {\bibfnamefont {M.-M.}\ \bibnamefont
  {B\'e}}, \bibinfo {author} {\bibfnamefont {V.}~\bibnamefont {Chist\'e}},
  \bibinfo {author} {\bibfnamefont {C.}~\bibnamefont {Dulieu}}, \bibinfo
  {author} {\bibfnamefont {X.}~\bibnamefont {Mougeot}}, \bibinfo {author}
  {\bibfnamefont {E.}~\bibnamefont {Browne}}, \bibinfo {author} {\bibfnamefont
  {V.}~\bibnamefont {Chechev}}, \bibinfo {author} {\bibfnamefont
  {N.}~\bibnamefont {Kuzmenko}}, \bibinfo {author} {\bibfnamefont
  {F.}~\bibnamefont {Kondev}}, \bibinfo {author} {\bibfnamefont
  {A.}~\bibnamefont {Luca}}, \bibinfo {author} {\bibfnamefont {M.}~\bibnamefont
  {Gal\'an}}, \bibinfo {author} {\bibfnamefont {A.}~\bibnamefont {Nichols}},
  \bibinfo {author} {\bibfnamefont {A.}~\bibnamefont {Arinc}},\ and\ \bibinfo
  {author} {\bibfnamefont {X.}~\bibnamefont {Huang}},\ }\href
  {http://www.bipm.org/utils/common/pdf/monographieRI/Monographie_BIPM-5_Tables_Vol5.pdf}
  {\emph {\bibinfo {title} {Table of Radionuclides}}},\ \bibinfo {series}
  {Monographie BIPM-5}, Vol.~\bibinfo {volume} {5}\ (\bibinfo  {publisher}
  {Bureau International des Poids et Mesures},\ \bibinfo {address} {Pavillon de
  Breteuil, F-92310 S\`evres, France},\ \bibinfo {year} {2010})\BibitemShut
  {NoStop}%
\bibitem [{\citenamefont {B\'e}\ \emph {et~al.}(2011)\citenamefont {B\'e},
  \citenamefont {Chist\'e}, \citenamefont {Dulieu}, \citenamefont {Mougeot},
  \citenamefont {Chechev}, \citenamefont {Kuzmenko}, \citenamefont {Kondev},
  \citenamefont {Luca}, \citenamefont {Gal\'an}, \citenamefont {Nichols},
  \citenamefont {Arinc}, \citenamefont {Pearce}, \citenamefont {Huang},\ and\
  \citenamefont {Wang}}]{TabRad_v6}%
  \BibitemOpen
  \bibfield  {author} {\bibinfo {author} {\bibfnamefont {M.-M.}\ \bibnamefont
  {B\'e}}, \bibinfo {author} {\bibfnamefont {V.}~\bibnamefont {Chist\'e}},
  \bibinfo {author} {\bibfnamefont {C.}~\bibnamefont {Dulieu}}, \bibinfo
  {author} {\bibfnamefont {X.}~\bibnamefont {Mougeot}}, \bibinfo {author}
  {\bibfnamefont {V.}~\bibnamefont {Chechev}}, \bibinfo {author} {\bibfnamefont
  {N.}~\bibnamefont {Kuzmenko}}, \bibinfo {author} {\bibfnamefont
  {F.}~\bibnamefont {Kondev}}, \bibinfo {author} {\bibfnamefont
  {A.}~\bibnamefont {Luca}}, \bibinfo {author} {\bibfnamefont {M.}~\bibnamefont
  {Gal\'an}}, \bibinfo {author} {\bibfnamefont {A.}~\bibnamefont {Nichols}},
  \bibinfo {author} {\bibfnamefont {A.}~\bibnamefont {Arinc}}, \bibinfo
  {author} {\bibfnamefont {A.}~\bibnamefont {Pearce}}, \bibinfo {author}
  {\bibfnamefont {X.}~\bibnamefont {Huang}},\ and\ \bibinfo {author}
  {\bibfnamefont {B.}~\bibnamefont {Wang}},\ }\href
  {http://www.bipm.org/utils/common/pdf/monographieRI/Monographie_BIPM-5_Tables_Vol6.pdf}
  {\emph {\bibinfo {title} {Table of Radionuclides}}},\ \bibinfo {series}
  {Monographie BIPM-5}, Vol.~\bibinfo {volume} {6}\ (\bibinfo  {publisher}
  {Bureau International des Poids et Mesures},\ \bibinfo {address} {Pavillon de
  Breteuil, F-92310 S\`evres, France},\ \bibinfo {year} {2011})\BibitemShut
  {NoStop}%
\bibitem [{\citenamefont {B\'e}\ \emph {et~al.}(2013)\citenamefont {B\'e},
  \citenamefont {Chist\'e}, \citenamefont {Dulieu}, \citenamefont {Mougeot},
  \citenamefont {Chechev}, \citenamefont {Kondev}, \citenamefont {Nichols},
  \citenamefont {Huang},\ and\ \citenamefont {Wang}}]{TabRad_v7}%
  \BibitemOpen
  \bibfield  {author} {\bibinfo {author} {\bibfnamefont {M.-M.}\ \bibnamefont
  {B\'e}}, \bibinfo {author} {\bibfnamefont {V.}~\bibnamefont {Chist\'e}},
  \bibinfo {author} {\bibfnamefont {C.}~\bibnamefont {Dulieu}}, \bibinfo
  {author} {\bibfnamefont {X.}~\bibnamefont {Mougeot}}, \bibinfo {author}
  {\bibfnamefont {V.}~\bibnamefont {Chechev}}, \bibinfo {author} {\bibfnamefont
  {F.}~\bibnamefont {Kondev}}, \bibinfo {author} {\bibfnamefont
  {A.}~\bibnamefont {Nichols}}, \bibinfo {author} {\bibfnamefont
  {X.}~\bibnamefont {Huang}},\ and\ \bibinfo {author} {\bibfnamefont
  {B.}~\bibnamefont {Wang}},\ }\href
  {http://www.bipm.org/utils/common/pdf/monographieRI/Monographie_BIPM-5_Tables_Vol7.pdf}
  {\emph {\bibinfo {title} {Table of Radionuclides}}},\ \bibinfo {series}
  {Monographie BIPM-5}, Vol.~\bibinfo {volume} {7}\ (\bibinfo  {publisher}
  {Bureau International des Poids et Mesures},\ \bibinfo {address} {Pavillon de
  Breteuil, F-92310 S\`evres, France},\ \bibinfo {year} {2013})\BibitemShut
  {NoStop}%
\bibitem [{\citenamefont {B\'e}\ \emph {et~al.}(2016)\citenamefont {B\'e},
  \citenamefont {Chist\'e}, \citenamefont {Dulieu}, \citenamefont {Kellett},
  \citenamefont {Mougeot}, \citenamefont {Arinc}, \citenamefont {Chechev},
  \citenamefont {Kuzmenko}, \citenamefont {Kib\'edi}, \citenamefont {Luca},\
  and\ \citenamefont {Nichols}}]{TabRad_v8}%
  \BibitemOpen
  \bibfield  {author} {\bibinfo {author} {\bibfnamefont {M.-M.}\ \bibnamefont
  {B\'e}}, \bibinfo {author} {\bibfnamefont {V.}~\bibnamefont {Chist\'e}},
  \bibinfo {author} {\bibfnamefont {C.}~\bibnamefont {Dulieu}}, \bibinfo
  {author} {\bibfnamefont {M.}~\bibnamefont {Kellett}}, \bibinfo {author}
  {\bibfnamefont {X.}~\bibnamefont {Mougeot}}, \bibinfo {author} {\bibfnamefont
  {A.}~\bibnamefont {Arinc}}, \bibinfo {author} {\bibfnamefont
  {V.}~\bibnamefont {Chechev}}, \bibinfo {author} {\bibfnamefont
  {N.}~\bibnamefont {Kuzmenko}}, \bibinfo {author} {\bibfnamefont
  {T.}~\bibnamefont {Kib\'edi}}, \bibinfo {author} {\bibfnamefont
  {A.}~\bibnamefont {Luca}},\ and\ \bibinfo {author} {\bibfnamefont
  {A.}~\bibnamefont {Nichols}},\ }\href
  {http://www.bipm.org/utils/common/pdf/monographieRI/Monographie_BIPM-5_Tables_Vol8.pdf}
  {\emph {\bibinfo {title} {Table of Radionuclides}}},\ \bibinfo {series}
  {Monographie BIPM-5}, Vol.~\bibinfo {volume} {8}\ (\bibinfo  {publisher}
  {Bureau International des Poids et Mesures},\ \bibinfo {address} {Pavillon de
  Breteuil, F-92310 S\`evres, France},\ \bibinfo {year} {2016})\BibitemShut
  {NoStop}%
\bibitem [{\citenamefont {Gillin}\ and\ \citenamefont
  {Middaugh}(2017)}]{DSCDimensions}%
  \BibitemOpen
  \bibfield  {author} {\bibinfo {author} {\bibfnamefont {K.}~\bibnamefont
  {Gillin}}\ and\ \bibinfo {author} {\bibfnamefont {E.}~\bibnamefont
  {Middaugh}},\ }\href
  {https://archive.opg.com/pdf_archive/Nuclear\%20Licencing\%20Documents/Pickering\%20Waste\%20Management\%20Licence\%20Renewal\%20(2017)/I011_92896-REP-01320-00003.pdf}
  {\emph {\bibinfo {title} {Pickering waste management facility safety
  assessment summary report}}},\ \bibinfo {type} {Tech. Rep.}\ (\bibinfo
  {institution} {Ontario Power Generation},\ \bibinfo {year}
  {2017})\BibitemShut {NoStop}%
\bibitem [{\citenamefont {Jonjev}(1996)}]{DryStoragePickering}%
  \BibitemOpen
  \bibfield  {author} {\bibinfo {author} {\bibfnamefont {S.}~\bibnamefont
  {Jonjev}},\ }\bibfield  {title} {\bibinfo {title} {Pickering dry storage},\
  }in\ \href
  {https://inis.iaea.org/collection/NCLCollectionStore/_Public/28/074/28074994.pdf}
  {\emph {\bibinfo {booktitle} {First International Conference On {CANDU} Fuel
  Handling Systems}}},\ \bibinfo {series and number} {\bibinfo {series} {24}\
  No.\ \bibinfo {number} {INIS-CA--0052}}\ (\bibinfo {year} {1996})\BibitemShut
  {NoStop}%
\bibitem [{\citenamefont {{IAEA}}(2022)}]{PickeringPower}%
  \BibitemOpen
  \bibfield  {author} {\bibinfo {author} {\bibnamefont {{IAEA}}},\ }\href@noop
  {} {\bibinfo {title} {{IAEA} power reactor information system page for
  pickering-1}} (\bibinfo {year} {2022}),\ \bibinfo {note}
  {\url{https://pris.iaea.org/PRIS/CountryStatistics/ReactorDetails.aspx?current=49},
  Last accessed on 2023-03-08}\BibitemShut {NoStop}%
\bibitem [{\citenamefont {Erlandson}(2023)}]{canduReactorSpectrum}%
  \BibitemOpen
  \bibfield  {author} {\bibinfo {author} {\bibfnamefont {A.}~\bibnamefont
  {Erlandson}},\ }\bibfield  {title} {\bibinfo {title} {Reactorgenic neutrino
  detection using liquid neon}\ }(\bibinfo  {publisher} {Presented at
  Magnificent CEvNS 2023},\ \bibinfo {year} {2023})\BibitemShut {NoStop}%
\bibitem [{\citenamefont {Strauss}\ \emph {et~al.}(2017)\citenamefont
  {Strauss}, \citenamefont {Rothe}, \citenamefont {Angloher}, \citenamefont
  {Bento}, \citenamefont {Gütlein}, \citenamefont {Hauff}, \citenamefont
  {Kluck}, \citenamefont {Mancuso}, \citenamefont {Oberauer}, \citenamefont
  {Petricca} \emph {et~al.}}]{Strauss_2017}%
  \BibitemOpen
  \bibfield  {author} {\bibinfo {author} {\bibfnamefont {R.}~\bibnamefont
  {Strauss}}, \bibinfo {author} {\bibfnamefont {J.}~\bibnamefont {Rothe}},
  \bibinfo {author} {\bibfnamefont {G.}~\bibnamefont {Angloher}}, \bibinfo
  {author} {\bibfnamefont {A.}~\bibnamefont {Bento}}, \bibinfo {author}
  {\bibfnamefont {A.}~\bibnamefont {Gütlein}}, \bibinfo {author}
  {\bibfnamefont {D.}~\bibnamefont {Hauff}}, \bibinfo {author} {\bibfnamefont
  {H.}~\bibnamefont {Kluck}}, \bibinfo {author} {\bibfnamefont
  {M.}~\bibnamefont {Mancuso}}, \bibinfo {author} {\bibfnamefont
  {L.}~\bibnamefont {Oberauer}}, \bibinfo {author} {\bibfnamefont
  {F.}~\bibnamefont {Petricca}}, \emph {et~al.},\ }\bibfield  {title} {\bibinfo
  {title} {Gram-scale cryogenic calorimeters for rare-event searches},\
  }\bibfield  {journal} {\bibinfo  {journal} {Phys. Rev. D}\ }\textbf {\bibinfo
  {volume} {96}},\ \href {https://doi.org/10.1103/physrevd.96.022009}
  {10.1103/physrevd.96.022009} (\bibinfo {year} {2017})\BibitemShut {NoStop}%
\bibitem [{\citenamefont {B.Cabrera}\ \emph {et~al.}(1985)\citenamefont
  {B.Cabrera}, \citenamefont {L.M.Krauss},\ and\ \citenamefont
  {F.Wilczek}}]{PhysRevLett.55.25}%
  \BibitemOpen
  \bibfield  {author} {\bibinfo {author} {\bibnamefont {B.Cabrera}}, \bibinfo
  {author} {\bibnamefont {L.M.Krauss}},\ and\ \bibinfo {author} {\bibnamefont
  {F.Wilczek}},\ }\bibfield  {title} {\bibinfo {title} {Bolometric detection of
  neutrinos},\ }\href {https://doi.org/10.1103/PhysRevLett.55.25} {\bibfield
  {journal} {\bibinfo  {journal} {Phys. Rev. Lett.}\ }\textbf {\bibinfo
  {volume} {55}},\ \bibinfo {pages} {25} (\bibinfo {year} {1985})}\BibitemShut
  {NoStop}%
\bibitem [{\citenamefont {Gottardi}\ \emph {et~al.}(2014)\citenamefont
  {Gottardi}, \citenamefont {Akamatsu}, \citenamefont {Barret}, \citenamefont
  {Bruijn}, \citenamefont {den Hartog}, \citenamefont {den Herder},
  \citenamefont {Hoevers}, \citenamefont {Kiviranta}, \citenamefont {van~der
  Kuur}, \citenamefont {van~der Linden} \emph {et~al.}}]{Gottardi_2014}%
  \BibitemOpen
  \bibfield  {author} {\bibinfo {author} {\bibfnamefont {L.}~\bibnamefont
  {Gottardi}}, \bibinfo {author} {\bibfnamefont {H.}~\bibnamefont {Akamatsu}},
  \bibinfo {author} {\bibfnamefont {D.}~\bibnamefont {Barret}}, \bibinfo
  {author} {\bibfnamefont {M.~P.}\ \bibnamefont {Bruijn}}, \bibinfo {author}
  {\bibfnamefont {R.~H.}\ \bibnamefont {den Hartog}}, \bibinfo {author}
  {\bibfnamefont {J.~W.}\ \bibnamefont {den Herder}}, \bibinfo {author}
  {\bibfnamefont {H.~F.~C.}\ \bibnamefont {Hoevers}}, \bibinfo {author}
  {\bibfnamefont {M.}~\bibnamefont {Kiviranta}}, \bibinfo {author}
  {\bibfnamefont {J.}~\bibnamefont {van~der Kuur}}, \bibinfo {author}
  {\bibfnamefont {A.~J.}\ \bibnamefont {van~der Linden}}, \emph {et~al.},\
  }\bibfield  {title} {\bibinfo {title} {{Development of {TES}-based detectors
  array for the X-ray Integral Field Unit (X-IFU) on the future x-ray
  observatory ATHENA}},\ }in\ \href {https://doi.org/10.1117/12.2056920} {\emph
  {\bibinfo {booktitle} {Space Telescopes and Instrumentation 2014: Ultraviolet
  to Gamma Ray}}},\ Vol.\ \bibinfo {volume} {9144},\ \bibinfo {organization}
  {International Society for Optics and Photonics}\ (\bibinfo  {publisher}
  {SPIE},\ \bibinfo {year} {2014})\ p.\ \bibinfo {pages} {91442M}\BibitemShut
  {NoStop}%
\bibitem [{\citenamefont {Augier}\ \emph {et~al.}(2021)\citenamefont {Augier}
  \emph {et~al.}}]{Ricochet:2021rjo}%
  \BibitemOpen
  \bibfield  {author} {\bibinfo {author} {\bibfnamefont {C.}~\bibnamefont
  {Augier}} \emph {et~al.} (\bibinfo {collaboration} {Ricochet}),\ }\bibfield
  {title} {\bibinfo {title} {{Ricochet Progress and Status}},\ }in\ \href@noop
  {} {\emph {\bibinfo {booktitle} {{19th International Workshop on Low
  Temperature Detectors}}}}\ (\bibinfo {year} {2021})\ \Eprint
  {https://arxiv.org/abs/2111.06745} {arXiv:2111.06745 [physics.ins-det]}
  \BibitemShut {NoStop}%
\end{thebibliography}
%

\end{document}